\documentclass[11pt]{article}

\sffamily

\usepackage{authblk}
\usepackage{hyperref} 
\usepackage{cite}
\usepackage{epsfig}
\usepackage{bm}
\usepackage{amssymb}
\usepackage{mathrsfs}
\usepackage{amsmath}
\usepackage{dsfont}
\usepackage{color}

\newcommand{\be}{\begin{equation}}
\newcommand{\ee}{\end{equation}}

\begin{document}

\title{\Large{Dual simulation of the 2d U(1) gauge Higgs model at topological angle $\theta = \pi\,$: Critical endpoint behavior}}

\vspace{25mm}

\author[1]{\vskip13mm 
Christof Gattringer\thanks{\href{mailto:christof.gattringer@uni-graz.at}{christof.gattringer@uni-graz.at}}}

\author[1]{Daniel G\"oschl\thanks{\href{mailto:daniel.goeschl@uni-graz.at}{daniel.goeschl@uni-graz.at}}}

\author[2]{Tin Sulejmanpasic\thanks{\href{mailto:tin.sulejmanpasic@gmail.com}{tin.sulejmanpasic@gmail.com}}\vspace{5mm}}

\affil[1]{\small University of Graz, Institute for Physics,  A-8010 Graz, Austria \vspace{3mm}}
\affil[2]{Philippe Meyer Institute, Physics Department, \'Ecole Normale Sup\'erieure,\newline 
PSL Research University, 24 rue Lhomond, F-75231 Paris Cedex 05, France}

\begin{titlepage}
\maketitle
\thispagestyle{empty}

\begin{abstract}
We simulate the 2d U(1) gauge Higgs model on the lattice with a topological angle $\theta$. The corresponding 
complex action problem is overcome by using a dual representation based on the Villain action appropriately 
endowed with a $\theta$-term. The Villain action is interpreted as a non-compact gauge theory whose center symmetry is 
gauged and has the advantage that the topological term is correctly quantized  so that $2\pi$ periodicity in $\theta$ is 
intact. Because of this the $\theta = \pi$ theory has an exact $\mathds{Z}_2$ charge-conjugation symmetry $C$, which 
is spontaneously broken when the mass-squared of the scalars is large and positive. Lowering the mass squared the 
symmetry becomes restored in a second order phase transition. Simulating the system at $\theta = \pi$ in its dual form 
we determine the corresponding critical endpoint as a function of the mass parameter. Using a finite size scaling analysis 
we determine the critical exponents and show that the transition is in the 2d Ising universality class, as expected. 
\end{abstract}

\end{titlepage}

\section{Introduction}
Perhaps one of the most interesting aspects of quantum field theories is that they allow for topological 
$\theta$-terms -- terms which couple to the global properties of the theory, rather than local degrees of freedom. As such 
they are invisible to perturbation theory, which is local in character. Still, the $\theta$-terms can significantly change the 
behavior of the theory, in particular since such terms typically violate some symmetries, so their inclusion has huge 
phenomenological consequences.

On the other hand the effect of such terms is difficult to study. Detailed analytic studies of their effects either require higher 
symmetry (e.g., supersymmetry or integrability) or a regime which is weakly coupled and amenable to computations. On 
the other hand such terms are imaginary even in the Euclidean formulation, such that first-principle numerical simulations 
are hindered by a complex-action problem.

The system we analyze in this paper is the U(1) gauge Higgs model in two dimensions with a topological term. Such 
models are tightly connected with various interesting systems. In condensed matter physics the U(1) gauge Higgs model 
with two scalar fields is an effective theory of $SU(2)$ anti-ferromagnetic spin chains, falling into two separate universality 
classes:   integer spin with $\theta=0$ and  half-integer spin for $\theta=\pi$. Moreover, in the large spin limit the 
antiferromagnetic spin chain becomes a $O(3)$ nonlinear sigma model\footnote{More accurately, because the target 
space is $S^2$, it should be referred to as the $S^2$ nonlinear sigma model.} with $\theta=2\pi S$, where $S$ is the spin 
quantum number of the underlying system. Finally, such models, being asymptotically free and having classical instanton 
solutions, are popular also in the particle physics community as toy models of QCD which shares some similar properties. 
The work we present here is a first step to studying such systems from first principles.

Two values of the topological angle, $\theta=0$ and $\theta = \pi$, are special  because in those cases the system is 
invariant under charge conjugation $C$, which is a global $\mathds{Z}_2$ symmetry. In the spin systems this symmetry 
corresponds to a translational symmetry $\!\mod 2$. Such a symmetry may break spontaneously or may be preserved in 
the ground-state. For $\theta=\pi$ a scenario that breaks $C$ spontaneously is ensured when the scalar fields are 
sufficiently massive, in which case the theory is a free $U(1)$ gauge theory and is exactly solvable\footnote{In this limit the 
system possesses an emergent $U(1)$ center symmetry, and there is a 't~Hooft anomaly between $C$ and the center 
symmetry which ensures that $C$ must be spontaneously broken. See the related work \cite{Gaiotto:2017yup,Gaiotto:2017tne,DiVecchia:2017xpu,Komargodski:2017dmc,Komargodski:2017smk,Tanizaki:2017qhf,Tanizaki:2018xto,Sulejmanpasic:2018upi}.}. On the other hand we expect that as the mass-squared of the scalars decreases 
towards negative infinity the system will undergo a transition into a trivially gapped phase\cite{Komargodski:2017dmc} (i.e., 
a phase with a gap, and no ground state degeneracy or topological order), which is conjectured to be a 2d Ising transition. 
This case is in distinction to the case with multiple scalar flavors in which case there is a 't Hooft anomaly between a global 
flavor symmetry and/or space-time symmetries and charge conjugation $C$ which forbids a trivially gapped phase. See 
\cite{Gaiotto:2017yup,Gaiotto:2017tne,DiVecchia:2017xpu,Komargodski:2017dmc,Komargodski:2017smk,Tanizaki:2017qhf,Tanizaki:2018xto,Sulejmanpasic:2018upi} for related discussions.

In the presence of a topological term Monte Carlo simulations are usually plagued by the complex action problem.
A possible way to solve the complex action problem is to exactly map the lattice model to a dual form in terms 
of worldlines and worldsheets, which is the approach we follow here 
(compare \cite{Endres:2006xu,Chandrasekharan:2008gp,Mercado:2013yta,Gattringer:2012jt,Korzec:2012fa,Mercado:2013ola,Korzec:2013dra,Kloiber:2014dfa,Schmidt:2015cva,Gattringer:2015baa,Bruckmann:2015sua,Bruckmann:2015hua} 
for use of these techniques in abelian gauge Higgs systems related to the model studied here). More specifically we 
use the dual formulation that is based on the Villain action \cite{Villain:1974ir} appropriately endowed with a $\theta$-term 
(for an earlier study of the same system with the Wilson action see \cite{Gattringer:2015baa}). 

We will arrive at the Villain action from considering a non-compact $U(1)$ lattice gauge theory coupled to matter. Such a 
theory has a non-compact $\mathbb Z$ center symmetry, and does not allow for a $\theta$-term (total fluxes on the closed 
manifold must be identically zero). By gauging the center symmetry, we must introduce a plaquette (a 2-form) 
$\mathbb Z$-valued gauge field, which allows the  global $\mathbb Z$ symmetry to be promoted to a gauge symmetry. 
Furthermore this allows for a natural introduction of a properly quantized $\theta$-term. We show that by choosing an 
appropriate gauge the action reduces to the Villain action which we use. However, note that there are other gauges which 
may be of interest in simulating fermionic theories whose dual descriptions are still plagued by a sign-problem.

Using the Villain action has the advantage that the dual form 
(a short derivation of the dual form for the Villain action is given in Section 2.2) implements the charge conjugation
symmetry that emerges at $\theta = \pi$ as an exact global $\mathds{Z}_2$ symmetry for the discrete dual 
variables -- we discuss this dual form of the symmetry in detail in Section 2.3.

We study the topological charge and the topological susceptibility numerically as a function of the mass parameter of the 
Higgs field, keeping the quartic coupling fixed. We find that the system undergoes the expected second order transition 
at a critical value of the mass parameter with the topological charge being the corresponding order parameter.  
We use finite size scaling techniques to show that the critical exponents are those of the 2d Ising model.

\section{The model and its dual representation}

\subsection{Continuum description and phase diagram}

A detailed discussion of the continuum expectations for the phase diagram can be found in 
\cite{Komargodski:2017dmc}, and here we only highlight some important points for clarity and completeness.

The problem can be formulated in the continuum with the action
\be
S=\int_{\mathds T^2} d^2x \left(|D_\mu\phi|^2+m^2 |\phi|^2+\lambda |\phi|^4+\frac{1}{2e^2}F_{12}^2+
\frac{i\theta}{2\pi} F_{12}\right)\;,
\ee
where $D_\mu=\partial_\mu+iA_\mu$ is the $U(1)$ covariant derivative, and $F_{12}=\partial_1A_2-\partial_2A_1$ is the field strength tensor. The theory is considered on a two-torus $\mathds T^2$. In that case the flux of $F_{12}$ is quantized in integer units of $2\pi$. For that reason the partition function,
\be
Z \; = \; \int \! D[\phi] \int \! D[A] \; e^{-S[\phi,A,\theta]} \; ,
\ee
is periodic with respect to the shifts $\theta\rightarrow \theta+2\pi$. 

We can analyze the system semi-classically in the regime where $|m^2|\gg e^2$.  
If $m^2$ is positive and large we can integrate out the $\phi$-field and end up with a pure gauge theory in 
1+1 dimensions, which can be exactly solved. The spectrum is given by
\be
E_k \; = \; \frac{e^2_{eff}}{2}\left(k-\frac{\theta}{2\pi}\right)^2L \; ,
\ee
where $L$ is the spatial volume. To one-loop the effective coupling $e_{eff}^2$ is related to the bare 
coupling $e^2$ as
\be
e_{eff}^2 \; = \; \frac{e^2}{1+\frac{e^2}{12\pi m^2}} \; ,
\ee 
which is valid as long as $m^2\gg e^2$.

\begin{figure}[tbp] 
   \centering
   \includegraphics[width=3.1in]{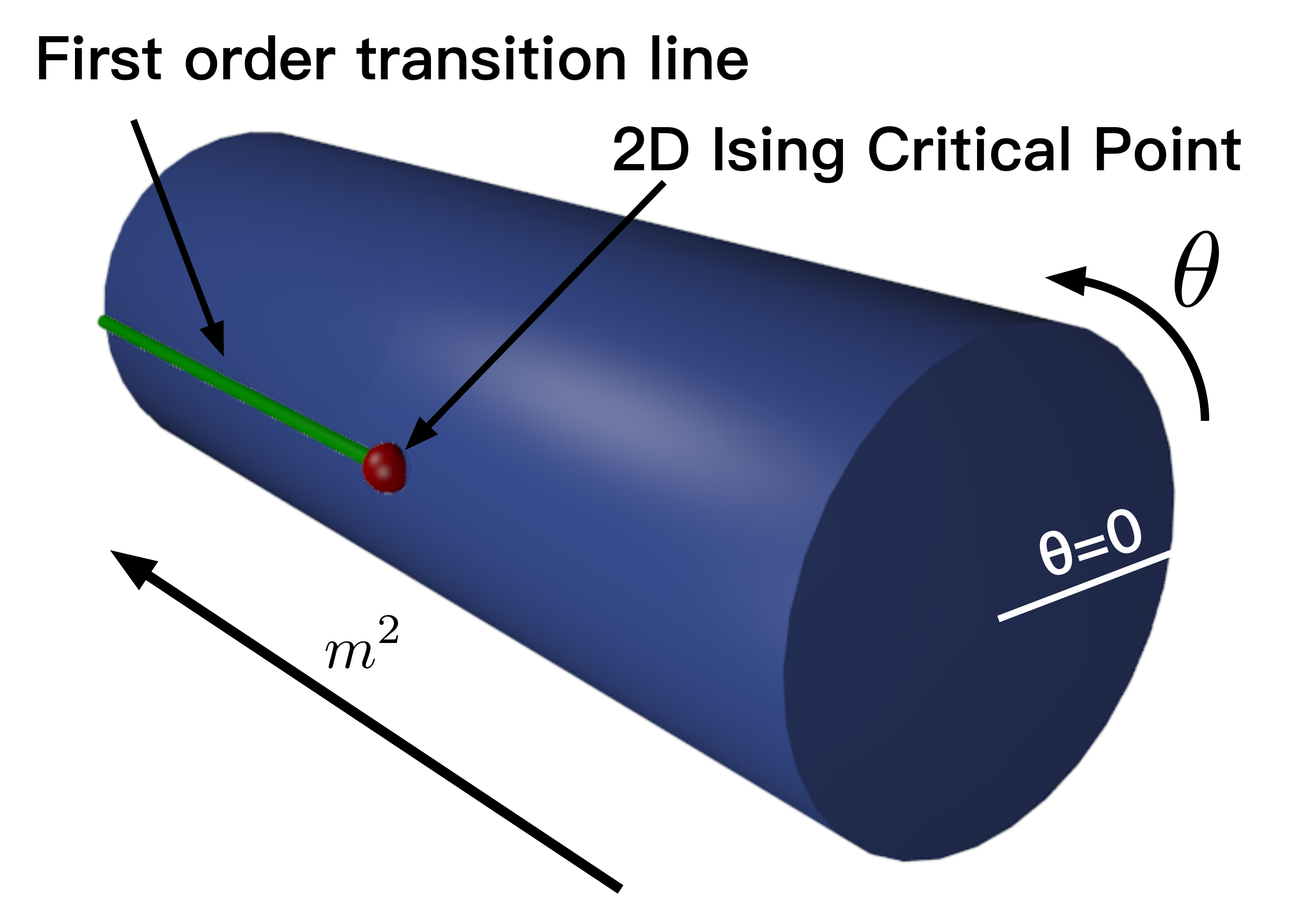} 
   \caption{The phase diagram of the theory as function of $m^2$ and $\theta$, as suggested in 
   \cite{Komargodski:2017dmc}.} 
   \label{fig:phase_diag}
\end{figure}

Notice that the spectrum is invariant under the change $\theta\rightarrow \theta+2\pi$, $k\rightarrow k+1$. Particularly 
interesting is the $\theta=\pi$ case, where the ground state is degenerate with equal energies for
$k = 0$ and $k=1$, i.e., $E_0 \, = \, E_1$. The topological susceptibility $\chi_t$ for both of these ground states 
is entirely determined by the coupling $e^2_{eff}$,
\be
\chi_t \; = \; -\frac{1}{L}\partial_\theta^2 E_{0,1} \; = \; -e^2_{eff}/(2\pi)^2 \; .
\ee

On the other hand we could also take $m^2\ll -e^2$. In this case the system is in the Higgs phase, 
with the modulus of the $\phi$-field weakly fluctuating. Let us set $\phi=e^{i\varphi}v$. 
We then have a term
\be
v^2 (\partial_\mu\varphi+A_\mu)^2
\ee
in the effective action. This term is minimized by  $A_\mu=-\partial_\mu\varphi$, i.e., a pure gauge. 
The vacuum is unique and charge conjugation invariant ($C$ invariant). Furthermore, the $\theta$-dependence is 
determined by instanton configurations whose size is cut off by the mass of the $\phi$ fluctuations which is 
proportional to $|m^2|$, allowing for instanton calculations. This yields the $\theta$-dependence of the ground 
state energy given by
\be
E_0(\theta) \; = \; -K e^{-S_0}\cos(\theta)L \; + \; ... \; ,
\ee
where $S_0$ is the instanton action and $K$ is a mass-dimension 2 pre-factor. 

As we change the parameter $m^2$ from large and positive to large and negative at $\theta=\pi$, we must 
encounter a $\mathds Z_2$ 2d Ising phase transition. The phase diagram as a function of $m^2$ and 
$\theta$ is expected to be as in 
Fig.~\ref{fig:phase_diag}. The bulk of the current work is to confirm this picture by first-principle simulations.

\subsection{Lattice representation: Gauging the center symmetry and the Villain action}

As we have mentioned several times already, our dualization starts from the Villain action. In order to be useful for our 
purposes we first have to properly define the $\theta$-term for the Villain action. Further, since we are specifically 
interested in the  physics at $\theta=\pi$, and in particular the realization of the charge conjugation symmetry at this point, 
it is vital that the topological charge is properly quantized.  To see how to do this, we will start from a slightly 
unconventional starting point and consider a non-compact gauge-theory, which in the presence of matter has a 
$\mathbb Z$ center symmetry. This symmetry is somewhat special, as it acts on line operators (the Polyakov loops) 
and is dubbed a 1-form symmetry (as distinct from the usual 0-form symmetry). 
This terminology follows \cite{Gaiotto:2014kfa}, and is rooted in the fact that 1-form symmetries act on line 
operators (i.e., integrals of 1-forms) as opposed to point 
operators, which can be thought of as 0-forms. To turn the theory into a compact gauge theory this symmetry has to be 
gauged with a plaquette (or 2-form) $\mathbb Z$-valued gauge field $B_x$. We will see that by doing this the theory has a 
natural $\theta$-term -- a sum over the plaquette based fields $B_x$. This theory is endowed with a 1-form discrete gauge 
symmetry in addition to the usual 0-form gauge symmetry. By fixing a gauge, the formulation becomes equivalent to the 
Villain action, endowed with a $\theta$-term.

In the conventional lattice representation the dynamical degrees of freedom for the 2-dimensional gauge-Higgs 
model are the complex valued field variables $\phi_x \in \mathds{C}$ assigned to the sites $x$ 
and the U(1) valued link variables $U_{x,\mu} \in$ U(1) assigned to the links $(x,\mu)$ 
of the lattice. Both variables obey periodic boundary conditions on the lattice of size $N_S\times N_T$ 
which we denote by $\Lambda$.
The corresponding partition sum is then given by 
\begin{equation}
Z \; = \; \int \! D[U] \; B_G[U] \; Z_H[U]  \; .	
\label{Zconventional}
\end{equation}
Here $B_G[U] = e^{-S_G[U]}$ is the Boltzmann factor for the gauge fields with action $S_G[U]$ and 
$Z_H [U] = \int \! D[\phi] \; e^{-S_H[\phi,U]}$ is the partition sum for the Higgs field in a 
background gauge field configuration $U$, where $S_H[\phi,U]$ is the Higgs action in the gauge background.

However, here we will take an approach different from the conventional lattice discretization, 
which, as we shall soon see, will result in the Villain action. 
To start let us consider a non-compact gauge theory on a lattice. Such a theory has $\mathds{R}$-valued link 
variables $A_{x,\mu}$, assigned to the links  $(x,\mu)$. We define the discretized field strength $F_x$ as
\begin{equation}
F_x \; = \;  A_{x+\hat{1},2} \,- \, A_{x,2} \, - \, A_{x+\hat{2},1}  \, + \, A_{x,1} \; =\; (dA)_x  \; .
\label{Fx}
\end{equation}
In the second step we have, for further use, also defined a lattice derivative $d$, acting on link fields.
The $F_x$ are assigned to the plaquettes of the lattice which we label by their lower left corner $x$. 
Now we define the partition function as
\be
Z \; = \; \left[ \prod_{x,\mu} \int_{-\infty}^\infty dA_{x,\mu} \right] e^{\, - \frac{\beta}{2} \sum_x F_x^2} \; 
Z_H\left[\{e^{iA_{x,\mu}}\}\right] \; ,
\ee
where we have discretized the gauge field action and defined the measure for the non-compact gauge fields as the 
product measure over all links. $Z_H$ is the same Higgs field partition sum as before, but in the argument we made 
explicit the representation of the gauge links as $U_{x,\mu} = e^{iA_{x,\mu}}$.

Note that as it stands the partition function is divergent and hence ill defined. The divergence comes about 
because of a global non-compact center symmetry, as well as the non-compact gauge group. The center symmetry 
can be thought of as a symmetry which shifts the link fields, $A_{x,\mu} \rightarrow A_{x,\mu} +\lambda_{x,\mu}$, 
where $\lambda_{x,\mu} \in 2 \pi \times\mathds Z$, with the constraint that $(d\lambda)_x=0 \; \forall x$. 
Such transformations are global $\mathds Z$-valued 1-form symmetry transformations as they shift  
Polyakov loops by a constant, $\sum_{(x,\mu) \in l}  A_{x,\mu}  \rightarrow 
\sum_{(x,\mu) \in l} A_{x,\mu} + \sum_{(x,\mu) \in l}  \lambda_{x,\mu}$, where $l$ denotes a closed non-contractible 
loop on our lattice. Notice that the shift of the loop is an integer multiple of $2\pi$ 
independent of its shape. In the absence of matter the link shifts $\lambda_{x,\mu}$ can also be chosen as sets of 
real numbers with $(d\lambda)_x=0 \; \forall x$.

The theory so far is a non-compact lattice gauge theory, and as such it does not have a $\theta$-term. 
It is also ill-defined because of the infinity associated with the integration measure.

To address these two issues, we will now gauge the $1$-form center 
symmetry\footnote{The term $1$-form symmetry was coined in 
\cite{Gaiotto:2014kfa} to distinguish the symmetries which act on line-operators (i.e., 1-form integrals) and which act on 
point operators (i.e., 0-forms).}. This promotes the set of $\lambda_{x,\mu}$ to be an arbitrary set of integer multiples of 
$2\pi$, without the constraints $(d\lambda)_x=0$. Clearly the field strength is not gauge 
invariant because now $F_x$ transforms as $F_x\rightarrow F_x+(d\lambda)_x$. To make it gauge invariant we must 
introduce a plaquette (2-form) gauge field, i.e., a $2\pi\mathds Z$-valued field $B_x$ on the plaquettes $x$, and replace 
$F_x \rightarrow F_x + B_x$. The transformation $A_{x,\mu} \rightarrow A_{x,\mu} +\lambda_{x,\mu}, \,
B_x\rightarrow B_x- (d\lambda)_x$ is a gauge symmetry that makes $F_x + B_x$ invariant.  Note that this gauge 
symmetry is \emph{in addition} to the usual gauge symmetry variation which sends $A_{x,\mu}\rightarrow A_{x,\mu}+
\phi_{x+\hat\mu}-\phi_{ x}$. One should not be surprised that we managed to formulate a theory which has more gauge 
symmetries than the usual formulation of the compact-gauge theories, as gauge symmetries are redundancies, and are 
not a measurable feature of the theory. Indeed one could always just fix the gauge, eliminating the redundancy altogether, 
and the physics should remain unchanged.

Notice that now also Wilson loops in the form $\sum_{(x,\mu) \in l} A_{x,\mu}$, 
where the loop $l$ closes trivially on the torus are not gauge invariant under the 
1-form gauge symmetry described above. This means they are not observables in our theory. Instead a Wilson loop 
$l$ defined as $e^{ik \sum_{(x,\mu) \in l} A_{x,\mu}}$ with $k\in\mathds Z$ is indeed invariant and thus an observable. 
This indicates that  by gauging the $\mathds Z$ center symmetry of the non-compact gauge theory, we have 
obtained a compact gauge theory, with the gauge action given by
\be
S_G \; = \; \frac{\beta}{2} \sum_x (F_x+B_x)^2 \; .
\ee
We note that the theory can be endowed with a natural $\theta$-term given by
\be
S_{\theta}\; = \; i\frac{\theta}{2\pi}\sum_x (F_x + B_x) \; = \; i\frac{\theta}{2\pi}\sum_x B_x \;.
\label{stheta}
\ee
Note that this is a natural $\theta$-term since it is related to $\int d^2x F_x$ in 
the continuum which was replaced by $\sum_x (F_x+B_x)$ in our discretization. On the lattice $F_x$ is a total 
derivative, so it drops out when summed over $x$, such that $\sum_x(F_x + B_x) = \sum_x B_x$, which is the form 
of $S_\theta$ on the rhs.~of (\ref{stheta}). Clearly $e^{-S_\theta}$ is gauge 
invariant and moreover, since $B_x \in 2\pi\mathds Z$, the partition function with a topological 
term based on (\ref{stheta})  is periodic under 
$\theta \rightarrow \theta+2\pi$. Note that at the special values $\theta=0$ and $\theta = \pi\mod 2\pi$ the system 
also has a charge conjugation symmetry $C$, implemented as $\phi_x\rightarrow \phi_x^*$, 
$A_{x,\mu} \rightarrow - A_{x,\mu}$ and $B_x \rightarrow -B_x$. This charge conjugation symmetry and our
correct implementation on the lattice will play a key role for the critical endpoint behavior discussed in this paper. 

Finally we address the question of ill-definitnes due to an infinity which stems from the integration range
$(-\infty,\infty)$ of the non-compact link fields $A_{x,\mu}$. Because we can always use a gauge symmetry
 $A_{x,\mu} \rightarrow A_{x,\mu} + \lambda_{x,\mu}$, where $\lambda_{x,\mu} = 2\pi \, m_{x,\mu}$ with 
 $m_{x,\mu} \in \mathds{Z}$, we can use the gauge invariance to fix the gauge such that 
 $A_{x,\mu} \in (-\pi,\pi]$. To impose this gauge let us introduce a function $\Theta_{-\pi,\pi}(x)$ which equals 1 for 
 $-\pi< x\le\pi$ and vanishes otherwise. By inserting unity in the form
\be
1 \; = \; \prod_{x,\mu} \sum_{k_{x,\mu} \in \mathds{Z} }\Theta_{-\pi,\pi}(A_{x,\mu}+2\pi k_{x,\mu}) \; ,
\ee
the partition function turns into
\begin{multline}
Z \; = \; \sum_{\{k\}} \sum_{\{B\}} \left[\prod_{x,\mu} \int_{-\infty}^\infty dA_{x,\mu}\; 
\Theta_{-\pi,\pi}(A_{x,\mu}+2\pi k_{x,\mu}) \right] \\ 
\times e^{-\frac{\beta}{2}\sum_x (F_x+B_x)^2 + i\frac{\theta}{2\pi}\sum_x B_x} \; Z_H(\{e^{iA_{x,\mu}}\}) \; ,
\end{multline}
where $\sum_{\{k\}} = \prod_{x,\mu} \sum_{k_{x,\mu} \in \mathds{Z}}$ and 
$\sum_{\{B\}} = \prod_{x} \sum_{B_x \in 2\pi \mathds{Z}}$.  We can now perform a gauge transformation 
$A_{x,\mu} \rightarrow A_{x,\mu} - 2\pi k_{x,\mu}$, which brings $Z$ into the form
\be
Z \; = \; \sum_{\{k\}} \sum_{\{B\}}
\left[ \prod_{x,\mu} \int_{-\pi}^\pi dA_{x,\mu} \right] e^{-\frac{\beta}{2} \sum_x (F_x+B_x)^2+
i\frac{\theta}{2\pi} \sum_x B_x } \; Z_H(\{e^{iA_{x,\mu}}\}) \;.
\ee
Now we note that the infinite sum $\sum_{\{k\}}$ is just a multiplicative factor and can be dropped. 
After dropping this term, and setting $B_x=2\pi n_x$, where $n_x\in \mathds Z$, we obtain the Villain 
representation \cite{Villain:1974ir} of the partition function, now containing the correctly discretized $2\pi$ periodic 
$\theta$-term that implements charge conjugation as an exact $\mathds{Z}_2$ symmetry.

It is important to stress that the Villain action is just a particular gauge in our discretization framework. We 
could have made other choices. For example we could make a choice which eliminates the field  $B_x$ from 
all plaquettes except one, 
labelled by $x_0$. What makes this gauge attractive is that we could then use it to propose ``instanton'' updates 
simply by proposing a change in $B_{x_0}$. This eliminates the topological charge sampling problem, 
because it reduces it to just one variable: the value of $B_{x_0}$. While here we will not make use of this gauge, 
it could potentially be of use when simulating theories with fermions, in which case dual descriptions are still 
plagued with the sign-problem.

We can now summarize the lattice discretization by providing the final form of the Boltzmann factor and 
specifying also the Higgs field action and Higgs field partition sum. The gauge field Boltzmann factor in the 
Villain form with the topological term is given by
\begin{eqnarray}
B_G[U] & = & \prod_{x\in\Lambda} \sum_ {n_x\in\mathds{Z} }
e^{\, - \, \frac{\beta}{2}\left(F_x+2\pi n_x\right)^2 \, - \, i \theta n_x }  
\nonumber \\
& = & \prod_{x\in\Lambda} \sum_ {n_x\in\mathds{Z} }
e^{\, - \, \frac{\beta}{2}\left(F_x+2\pi n_x\right)^2 \, - \, i \frac{\theta}{2\pi} (F_x + 2\pi n_x) } \; ,
\label{B_G_2}
\end{eqnarray}
where in the second expression for $B_G[U]$ we have inserted $F_x$ in the exponent coupling to $\theta$, 
which, as we have already discussed, is an equivalent form because of $\sum_x F_x = 0$. The latter 
expression will be used in the appendix for transforming $B_G[U]$ to its dual form. 

The partition function $Z_H[U]$ for the Higgs field in a background $U = \{ e^{iA_{x,\mu}}\}$ reads 
\begin{equation}
Z_H[U] \; = \; \int D[\phi] \ e^{-S_H[\phi,U]} \; ,
\label{ZH_def}
\end{equation}
where the action $S_H[\phi,U]$ for the Higgs field is given by 
\begin{equation}
S_H[\phi,U] \; = \; \sum_{x\in\Lambda}\left[ M |\phi_x|^2 + \lambda |\phi_x|^4 - \sum_{\mu=1}^{2} 
\Big(\phi_{x}^{\ast} U_{x,\mu} \phi_{x+\hat{\mu}}+c.c.\Big) \right] \; .
\label{S_Higgs}
\end{equation}
$\lambda$ denotes the quartic coupling and the mass parameter $M$ is defined as $M = 4 + m^2$,  
where $m$ is the bare mass. 

The path integral measures are a product of U(1) Haar measures for the gauge fields and a product 
of integrals over $\mathds{C}$ for the Higgs field,
\begin{equation}
	\int D[U] \; = \; \prod_{x,\mu} \int_{-\pi}^{\pi} \frac{dA_{x,\mu}}{2\pi} \quad , \quad
	\int D[\phi] \; = \; \prod_x \int_{\mathds{C}}  \frac{d\phi_x}{2\pi} \;  .
\label{measures}	
\end{equation}

The charge conjugation transformation which we have addressed corresponds to
$A_{x,\mu} \rightarrow - A_{x,\mu}, \; \phi_x \rightarrow \phi_x^\star$. It is easy to see that this transformation
is a symmetry of the action $S_H[\phi,U]$ for the Higgs field (\ref{S_Higgs}) and of the path integral measure.
For the field strength the 
transformation gives rise to $F_x \rightarrow - F_x$ such that $n_x \rightarrow - n_x$ compensates this 
in the quadratic part of the exponent of the gauge field Boltzmann factor (\ref{B_G_2}), while the linear part
of the exponent is symmetric only at $\theta = \pi$ (and of course at the trivial point $\theta = 0$). 
Analyzing the spontaneous breaking of this symmetry for the $\theta = \pi$ case with Monte Carlo simulation 
techniques is the goal of this paper.

It is obvious that a non-zero vacuum angle in the representation (\ref{B_G_2}) gives rise to a complex Boltzmann 
factor and thus to a so-called complex action problem. This complex action problem can be solved completely by mapping
the partition sum exactly to a dual form where the new degrees of freedom are worldlines for matter fields and 
world-sheets for the gauge fields.

\subsection{Dual representation}

This section summarizes the exact transformation of the model in Villain formulation to the corresponding 
dual formulation in terms of worldlines and plaquette occupation numbers.
We begin the presentation of the mapping to the dual form for the Villain action by 
quoting from the literature the dual representation of the partition
function $Z_H[U]$ for the Higgs field in a gauge field background. This form is derived by expanding the 
Boltzmann factors for the individual nearest neighbor terms and subsequently integrating out the original 
field variables $\phi_x$ (see, e.g., the appendix of \cite{Mercado:2013yta} 
for the derivation of the form used here). In the dual form the new 
degrees of freedom are flux variables $j_{x,\mu}\in\mathds{Z}$ 
assigned to the links of the lattice and auxiliary variables $h_{x,\mu}\in\mathds{N}_0$ also living on the 
links. The partition sum $Z_H[U]$ is a sum 
$\sum_{\{j,h\}} \equiv \prod_{x,\mu} \sum_{j_{x,\mu}\in\mathds {Z}} \sum_{h_{x,\mu}\in\mathds{N}_0}$ 
over all possible configurations of these variables. It is given by
\begin{equation}
Z_H[U] \; = \; \sum_{\{j,h\}} W_H[j,h] \, \prod_x \delta\left(\vec{\nabla}\cdot \vec{j}_x\right) 
\, \prod_{x,\mu} \, \Big( U_{x,\mu} \Big)^{\; j_{x,\mu}} \; .
\label{ZHflux}
\end{equation}
Each configuration of the flux variables $j_{x,\mu}$ and the auxiliary variables 
$h_{x,\mu}$ comes with a weight factor $W_H[j,h]$
given by  
\begin{eqnarray}
W_H[j,h] &\!\!\!=\!\! \!& \left[\prod_{x,\mu} \frac{1}{\left(|j_{x,\mu}|\!+\!h_{x,\mu}\right)! \ h_{x,\mu}!} \right]
\left[\prod_x \! I\left(f_x\right)\right] , 
\label{W_H} \\ 
I\!\left(f_x\right)  &\!\!\!=\!\! \!&  \! \int_{0}^{\infty} \!\!\!\!\! dr \ r^{f_x+1} e^{-M r^2 -\lambda r^4}  \; ,
\label{If} \\
f_x &\!\!\! \equiv \!\! \!&\ \sum_{\mu}
\left[ |j_{x,\mu}| + |j_{x-\hat{\mu},\mu}| + 2\left( h_{x,\mu} + h_{x-\hat{\mu},\mu} \right) \right] \; .
\label{fx}
\end{eqnarray}
The first part of the weight factor, i.e., the product over the links collects the combinatorial factor from the expansion
of the nearest neighbor Boltzmann factors. This is followed by a product over all sites where the integrals 
$I\left(f_x\right)$ emerge when integrating out the moduli of the complex fields $\phi_x$. These integrals depend on 
the non-negative combinations $f_x$ of the flux and auxiliary variables given in (\ref{fx}). For the Monte Carlo 
simulation the integrals $I\left(f_x\right)$ are numerically pre-calculated and stored for a sufficient number of values 
$f_x$ for the chosen parameters $M$ and $\lambda$. 

Integrating over the complex phases of the fields $\phi_x$ generates constraints for the flux variables $j_{x,\mu}$.  
These constraints appear as a product of Kronecker deltas in (\ref{ZHflux}) where we use the notation 
$\delta(n) = \delta_{n,0}$. The constraints enforce vanishing divergence of the flux $j_{x,\mu}$ at every site 
$x$, i.e., $\vec{\nabla}\cdot \vec{j}_x \equiv \sum_{\mu} [ j_{x,\mu} - j_{x-\hat{\mu},\mu}] = 0 \; \forall x$.
As a consequence of the vanishing divergence constraint the admissible configurations of the flux variables 
$j_{x,\mu}$ have the form of closed loops of $j$-flux. The auxiliary variables $h_{x,\mu}$ are unconstrained.

Along the closed loops of the $j$-flux the U(1) variables on the links $(x,\mu)$ of the loop are activated 
according to the flux $j_{x,\mu}$. This gives rise to the last product in (\ref{ZHflux}) where for negative values of 
$j_{x,\mu}$ we have $\big( U_{x,\mu} \big)^{\; j_{x,\mu}} = \big( U_{x,\mu} \big)^{\;-| j_{x,\mu} |} = 
\big( U_{x,\mu}^\star \big)^{\; | j_{x,\mu}|}$, i.e., for links that are run through in negative direction the complex
conjugate link variable is taken into account.

The link variables along the loops now have to be integrated over with the gauge field Boltzmann factor $B_G[U]$. 
As we show in the appendix, one may use Fourier transformation with respect to the field strength $F_x$ to write
the Boltzmann factor of the Villain action in the form 
\begin{equation}
B_G[U] \; = \; \left[ \prod_x \sum_{p_x \in \mathds{Z}} \right] 
\, \left[ \prod_x \frac{1}{\sqrt{2\pi \beta}} \; 
e^{\, - \frac{1}{2 \beta} \big(p_x + \frac{\theta}{2\pi}\big)^2} \right]  \, \left[ \prod_x e^{i F_x \, p_x} \right] \; .
\label{B_G_dual}
\end{equation} 
At every plaquette of the lattice we sum over a plaquette occupation number $p_x \in \mathds{Z}$, and the corresponding
sum over all configurations will be denoted as $\sum_{\{p\}} \equiv  \prod_x \sum_{p_x \in \mathds{Z}}$. 
The configurations of the plaquette occupation numbers come with a Gaussian weight factor at every site that 
depends on both, the inverse gauge coupling $\beta$ and the vacuum angle $\theta$. For this weight factor we introduce
the abbreviation ($V = N_S N_T$ denotes the number of sites) 
\begin{equation}
W_G[p] \; = \; \left( \frac{1}{2\pi \beta} \right)^{\frac{V}{2}} 
\prod_x e^{\, - \frac{1}{2 \beta} \big(p_x + \frac{\theta}{2\pi}\big)^2} \; .
\label{W_G}
\end{equation}
The gauge field dependence of the Boltzmann factor $B_G[U]$ is collected in the last term of (\ref{B_G_dual}), 
which is a product over 
the Fourier factors $e^{i F_x p_x}$ at all plaquettes. Inserting the explicit expression (\ref{Fx}) we can reorganize this  
product,
\begin{eqnarray}
\prod_x e^{i F_x \, p_x} & = & 
\prod_x e^{ i \big[ A_{x+\hat{1},2} \, -  \, A_{x,2} \, - \, A_{x+\hat{2},1} \, + \, A_{x,1} \big] \, p_x}
\nonumber \\
& = &  \prod_x  e^{ i  A_{x,1} \big (p_x \, - \, p_{x-\hat{2}}\big)} \; e^{ i  A_{x,2} \big(p_{x-\hat{1}} \, - \, p_x\big)} \; ,
\end{eqnarray}
where in the last step we have shifted the nearest neighbor displacement from the $A_{x,\mu}$ to the plaquette
occupation numbers $p_x$ by accordingly relabelling the variable $x$ in the product over all sites. In this form we 
can combine this factor with the gauge field dependence of $Z_H[U]$ in (\ref{ZHflux}) and integrate over the link variables. 
We find
\begin{eqnarray}
&& \int D[U] \, \prod_{x,\mu} \, \Big( U_{x,\mu} \Big)^{\; j_{x,\mu}} \, \prod_x e^{i F_x \, p_x} 
\nonumber \\
&& = 
\left[\prod_{x,\mu} \int_{-\pi}^{\pi} \frac{dA_{x,\mu}}{2\pi}\right] \! 
\left[ \prod_{x,\mu} e^{\, i A_{x,\mu} \, j_{x,\mu}} \right] \!
\left[ \prod_x e^{ i  A_{x,1} \big (p_x \, - \, p_{x-\hat{2}}\big)} \; e^{ i  A_{x,2} \big(p_{x-\hat{1}} \, - \, p_x\big)} \right]
\nonumber \\
&& = \prod_x \delta\left(j_{x,1}+p_{x}-p_{x-\hat{2}}\right) \, \delta\left(j_{x,2}-p_{x}+p_{x-\hat{1}}\right) \; ,
\end{eqnarray}
where in the last step we have used $\int_{-\pi}^\pi d A/2\pi \; e^{\, iA \; n} = \delta(n)$. 
Thus the integration over the U(1) link 
variables gives rise to a second set of constraints which connect the $j$-flux and the plaquette occupation numbers $p_x$ 
such that at each link the sum of $j$-flux and flux from the plaquettes containing that link has to vanish.  

Putting things together we find the final form of the dual representation
\begin{equation}
Z =  \!\! \sum_{\{j,h,p\}} \! \! W_H[j,h] \, W_G[p] 
\prod_x \! \delta\!\left(\!\vec{\nabla}\cdot \vec{j}_x\! \right) \!
\delta\!\left(j_{x,1}+p_{x}-p_{x-\hat{2}}\right) \! \delta\!\left(j_{x,2}-p_{x}+p_{x-\hat{1}}\right) .
\label{dualZ}
\end{equation}
The partition function is a sum over all configurations of the link-based flux variables $j_{x,\mu} \in \mathds{Z}$, 
the link-based auxiliary variables $h_{x,\mu} \in \mathds{N}_0$ and the plaquette
occupation numbers $p_{x} \in \mathds{Z}$. The  $j_{x,\mu}$ must obey the zero divergence constraint that 
forces admissible configurations to have the form of closed loops of $j$-flux. A second set of constraints links these
fluxes to the plaquette occupation numbers such that the combination of $j$-flux and flux from the plaquettes
vanishes at each link of the lattice. As a consequence admissible configurations of the plaquette variables are either
closed surfaces formed by occupied plaquettes or open surfaces with $j$-flux at the boundaries. 

The configurations come with a weight factor $W_H[j,h]$ for the $j$-flux and the unconstrained auxiliary 
variables $h_{x,\mu}$ which together describe the Higgs field dynamics in the dual language. This weight 
factor is given explicitly in (\ref{W_H}). The second weight factor $W_G[p]$ for the plaquette occupation numbers 
governs the gauge dynamics and is given in (\ref{W_G}). The weight has the form of a Gaussian for the plaquette 
occupation number $p_x$ at each site. The width is given by the square root of the inverse gauge coupling and the 
mean of the Gaussian is at $\theta/2\pi$.  Obviously all weights are positive at arbitrary values of $\theta$ and thus
the dual representation solves the complex action problem. 

\subsection{Charge conjugation symmetry at $\theta = \pi$ and observables}

Let us now discuss an important aspect of the dual representation for the Villain action, i.e., the implementation 
of the charge conjugation symmetry at $\theta = \pi$  as global $\mathds{Z}_2$ symmetry for the discrete 
dual variables. At $\theta = \pi$ the product of Gaussians in the weight  
$W_G[p]$ (\ref{W_G}) is $\prod_x \exp \big( \!-\! \frac{1}{2 \beta} (p_x + 1/2)^2 \big)$. It is obvious that
the transformation 
\begin{equation}
p_x \; \longrightarrow \; - \,p_x \, - \, 1 \; \; \forall x \; ,
\label{transform1}
\end{equation} 
leaves the weight factor $W_G[p]$ invariant. In order to show that this gives rise to a full symmetry we 
need to discuss the interaction with the flux variables $j_{x,\mu}$, encoded in the constraints
$ \delta\!\left(j_{x,1}+p_{x}-p_{x-\hat{2}}\right) \! \delta\!\left(j_{x,2}-p_{x}+p_{x-\hat{1}}\right)$ appearing 
in (\ref{dualZ}). First we note that a uniform constant shift of all $p_x$, such as the additive term $- 1$ on the rhs.\ of 
(\ref{transform1}) leaves the constraints intact, since in each of the Kronecker deltas the difference of two plaquette
occupation numbers appears, such that a constant shift cancels. Flipping the sign of the $p_x$, which is the 
second ingredient of the transformation (\ref{transform1}), can be compensated in the constraints by changing the
signs of all $j$-fluxes,
\begin{equation}
j_{x,\mu} \; \longrightarrow \; - \, j_{x,\mu} \; \; \; \forall x, \mu \; .
\label{transform2}
\end{equation}
The final step for completely establishing the symmetry is to note that the flux variables $j_{x,\mu}$ enter in the
weight factors $W_H[j,a]$ in Eq.~(\ref{W_H}) only via their absolute values $|j_{x,\mu}|$, such that also the weight factor 
$W_H[j,a]$ remains unchanged. Thus the combination of (\ref{transform1}) and (\ref{transform2}) implements a 
global $\mathds{Z}_2$ symmetry for the discrete dual variables. The fact that the symmetry is $\mathds{Z}_2$ is 
obvious, since applying the transformation (\ref{transform1}), (\ref{transform2}) twice gives the identity transformation 
($p_x^\prime = p_x \;\; \forall x, \; j_{x,\mu}^\prime = j_{x,\mu} \; \; \forall x, \mu$).

In this paper we want to study the $\mathds{Z}_2$ symmetry and thus need a corresponding
order parameter. A suitable order parameter is given by the topological charge density\footnote{Below we define $q$ 
in such a way that it is not really the topological charge density, but differs from it by a multiplicative factor of $i$. 
We will still refer to it as topological charge density for brevity.} $q$. Its vacuum expectation
value $\langle q \rangle$ and the corresponding topological susceptibility $\chi_t$ are defined as
\begin{equation}
\langle q \rangle \; = \; - \frac{1}{V} \frac{\partial}{\partial\theta} \ln Z 	
\; , \qquad 
\chi_t \; = \; \frac{1}{V} \frac{\partial^2}{\partial\theta^2} \ln Z \; .
\end{equation}
As an additional observable we also study the gauge field action density $\langle s_G \rangle$ defined as
\begin{equation}
\langle s_G \rangle \; = \; - \frac{1}{V} \frac{\partial}{\partial\beta} \ln Z  \; .
\end{equation}
In order to determine the dual form of these observables we perform the derivatives with respect to $\theta$ 
using the dual representation for $Z$ and obtain 
\begin{align}
& \langle q \rangle  = \frac{1}{V}
\left \langle \frac{1}{2\pi\beta}\sum_x \! \left[  p_x + \frac{\theta}{2\pi} \! \right]\! \right\rangle \; , 
\label{qdual}
\\
& \chi_t= \frac{1}{V} \! \left[ \left \langle \!\! \left(\!\frac{1}{2\pi\beta} \! \sum_x \! 
\left[  p_x + \frac{\theta}{2\pi} \! \right]\! \right)^{\!\!\!2}\right\rangle - 
\left \langle \! \frac{1}{2\pi\beta}\!\sum_x \! \left[  p_x + \frac{\theta}{2\pi} \! \right]\! \right\rangle^{\!\!\!2} \right]
\! - \frac{1}{4 \pi^2 \beta} \, , \\
&\langle s_G \rangle  =  \frac{1}{2 \beta V} \!
\left \langle \sum_x \! \left[  1 - \frac{\left(p_x + \frac{\theta}{2\pi} \right)^2}{\beta} \right]\! \right\rangle \; .
\end{align}

It is easy to see that for $\theta = \pi$ the topological charge density $q$, which in the dual representation 
(\ref{qdual}) is given by  $q \, |_{\theta = \pi} = (V 2 \pi \beta)^{-1} \sum_x (p_x + 1/2)$, 
changes sign under the $\mathds{Z}_2$ transformation 
(\ref{transform1}), (\ref{transform2}), and thus is a suitable order parameter for the breaking of that symmetry. 
Note that on a finite lattice the $\mathds{Z}_2$ symmetry at $\theta = \pi$ cannot be broken spontaneously such that 
$\langle q \rangle$ vanishes exactly. Thus in our numerical analysis at $\theta = \pi$ we will partly 
evaluate $\langle | q | \rangle$, which also can be used to identify and analyze the phase transition. 

The situation which we find for the U(1) gauge-Higgs model at $\theta = \pi$ thus is reminiscent of a ferromagnet. 
The corresponding order parameter, i.e., the magnetization here is replaced by the topological charge density $q$. 
The role of the external magnetic field is played by 
\begin{equation}
\Delta \theta \; = \; \theta \, - \, \pi \; ,
\end{equation}
i.e., the topological angle shifted such that $\Delta \theta = 0$ corresponds to the symmetrical point where explicit 
symmetry breaking is absent. Finally we need to identify the control parameter that drives the system 
through the phase transition. For the ferromagnet this parameter is the inverse temperature, 
which here is replaced by the parameters $M$ and $\lambda$. In our analysis we will keep $\lambda$ fixed and use 
the mass parameter $M = 4 + m^2$ as the control parameter to drive the system at the symmetrical value 
$\Delta \theta = 0$ (i.e., $\theta = \pi$) through the expected phase transition.  

\section{Numerical simulation}

\subsection{Details of the simulation}

For the Monte Carlo updates of abelian gauge Higgs models in the dual representation different strategies 
can be followed. One type of strategy offers only trial configurations that already obey the constraints. This strategy 
can, e.g., be implemented by offering the following local changes
\begin{eqnarray} 
p_{x_0} \rightarrow p_{x_0} + \Delta , \quad 
&& j_{x_0,1} \rightarrow j_{x_0,1} - \Delta, \; 
j_{x_0 + \hat{1},2} \rightarrow j_{x_0 + \hat{1},2} - \Delta \; ,
\nonumber \\
&& j_{x_0 + \hat{2},1}  \rightarrow j_{x_0 + \hat{2},1} + \Delta, \; 
j_{x_0,2} \rightarrow j_{x_0,2} + \Delta\; ,
\label{localMC}
\end{eqnarray}
where $x_0$ is some site of the lattice and the increment $\Delta$ is randomly chosen in $\{-1, +1\}$. The proposed 
changes (\ref{localMC}) are accepted with the usual Metropolis probability that can easily be computed from the 
weights of the dual representation and the new configuration generated by (\ref{localMC}) obeys all constraints. 
Together with a conventional Metropolis update of the auxiliary variables $h_{x,\mu}$, which are not subject to constraints, 
the update (\ref{localMC}) already constitutes an ergodic algorithm. However, for faster decorrelation it is useful to also
add a global update of all plaquette occupation numbers,
\begin{equation} 
p_{x} \rightarrow p_{x} + \Delta  \quad \forall \; x \; ,
\label{blanket}
\end{equation}
where again $\Delta$ is an increment randomly chosen in $\{-1, +1\}$. 

The other strategy for the update of the dual gauge Higgs model is based on the idea of the worm algorithm 
\cite{Prokofev:2001ddj}. 
In an initial step one locally violates the constraints by changing some of the dual variables. Subsequent update steps 
randomly propagate the defect on the lattice until the starting point is reached and the defect is healed. All intermediate 
propagation steps are accepted with suitable Metropolis decisions and after the worm has closed, one again has a 
configuration that obeys all constraints and the total transition probability fulfills detailed balance. For applying 
the strategy to dual abelian gauge theories the structures generated in the propagation steps are surfaces of 
plaquette occupation numbers bounded by $j$-flux. This so-called surface worm strategy was discussed in detail in 
\cite{Mercado:2013yta} and in some parameter regions clearly outperforms the local updates in four dimensions. 
However, here we simulate the considerably less demanding 2-dimensional case where the updates 
(\ref{localMC}) and (\ref{blanket}) are sufficient for very accurate results.

In our numerical simulation we combine sweeps of the local updates (\ref{localMC}) and the global plaquette occupation 
number updates (\ref{blanket}). More specifically, for each parameter set we perform $10^5$ mixed sweeps for 
equilibration. The statistics ranges between $10^4$ and $10^7$ measurements, depending on the parameter region, 
i.e., the distance to the critical point. We typically use 20 to 50 (for the larger lattices) mixed sweeps in 
between measurements for decorrelation. The errors we show for our results are statistical errors determined 
from a jackknife analysis.

We carefully tested our dual simulation algorithms using several test cases. For pure gauge theory at arbitrary $\theta$
one can compute the partition sum in closed form: In that case the constraints for the remaining dual variables 
$p_x$ imply that only constant configurations $\forall x\!:  p_x = j \in \mathds{Z}$ are admissible and the partition 
function and observables are obtained as fast converging sums over $j \in \mathds{Z}$ (in fact $Z$ is given 
by the $\theta_3$ elliptic function). Our simulation algorithm reproduces this analytic result with very high statistics. 
A second case where we tested the dual simulation is the full theory at  $\theta = 0$. In that case the complex
action problem is absent and reference simulations with the conventional representation are possible. Also for this test
we found excellent agreement of results from the dual simulation with those from the conventional representation. 

In all our numerical simulations we kept the quartic coupling fixed at a value of $\lambda = 0.5$. For the inverse gauge 
coupling $\beta$ we used $\beta = 3.0$, and $\beta = 5.0$. 

\subsection{Results for bulk observables}

\begin{figure}[t!]
\centering      
\hspace*{-3mm}
\includegraphics[width=129mm,clip]{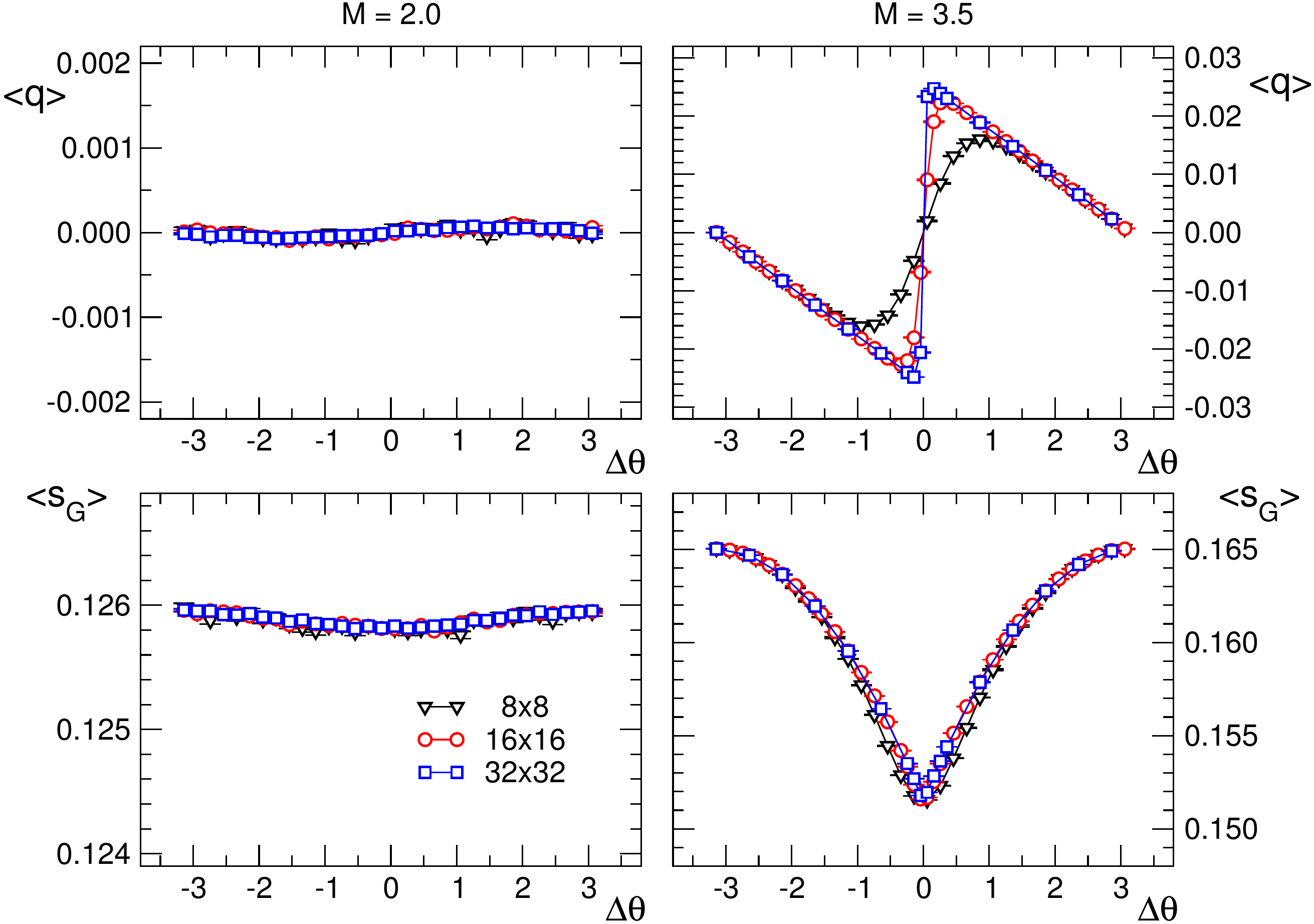}
\caption{The topological charge density $\langle q \rangle$ (top row of plots) and the gauge
action density  $\langle s_G \rangle$ (bottom) as a function of $\Delta \theta$. 
We compare different volumes and use $\beta = 3.0, \lambda = 0.5$. For the mass parameter $M = 4 + m^2$
we use $M = 2.0$ (lhs.\ column of plots) and $M = 3.5$ (rhs.).}
\label{S_Q_vs_theta}
\end{figure}

We begin the presentation of our results with a plot of the topological charge density $\langle q \rangle$ 
and the gauge action density $\langle s_G \rangle$ as a function of the symmetry breaking parameter $\Delta \theta$. 
We study the behavior for two different values of the  mass parameter $M = 4 + m^2$, given by $M = 2.0$ (lhs.\ plots) 
and $M = 3.5$ (rhs.). The other parameters are $\beta = 3.0$ and $\lambda = 0.5$ and in Fig.~\ref{S_Q_vs_theta} 
we show our results for three different volumes. 

It is obvious that for a mass parameter value of $M = 2.0$ both observables $\langle q \rangle$ 
and $\langle s_G \rangle$ show only a very weak dependence on the symmetry breaking parameter 
$\Delta \theta = \theta - \pi$ and essentially no scaling with the volume.  This clearly is different for 
$M = 3.5$, where in the thermodynamic limit we see the emergence of a first order jump in 
$\langle q \rangle$, which has also been observed in both, pure U(1) gauge 
theory in 2-dimensions \cite{Wiese:1988qz}, as well as in the simulation of the U(1) gauge Higgs system with the 
Wilson gauge action \cite{Gattringer:2015baa}. Also  the gauge action density is sensitive to $\Delta \theta$ at 
$M = 3.5$ and in the thermodynamic limit develops a cusp at the symmetrical point  $\Delta \theta = 0$,
i.e., at $\theta = \pi$.

From the fact that at $M = 3.5$ we observe a $\theta$-dependence of the observables with an emerging 
first order behavior at $\theta = \pi$ and only very weak dependence on $\theta$ for $M = 2.0$, 
we expect that the first order line at $\theta = \pi$ must have an endpoint at some critical value $M_c$ 
somewhere between $M = 2.0$ and $M = 3.5$.
 
\begin{figure}[t!]
\centering
\vspace*{-3mm}
\includegraphics[width=125mm,clip]{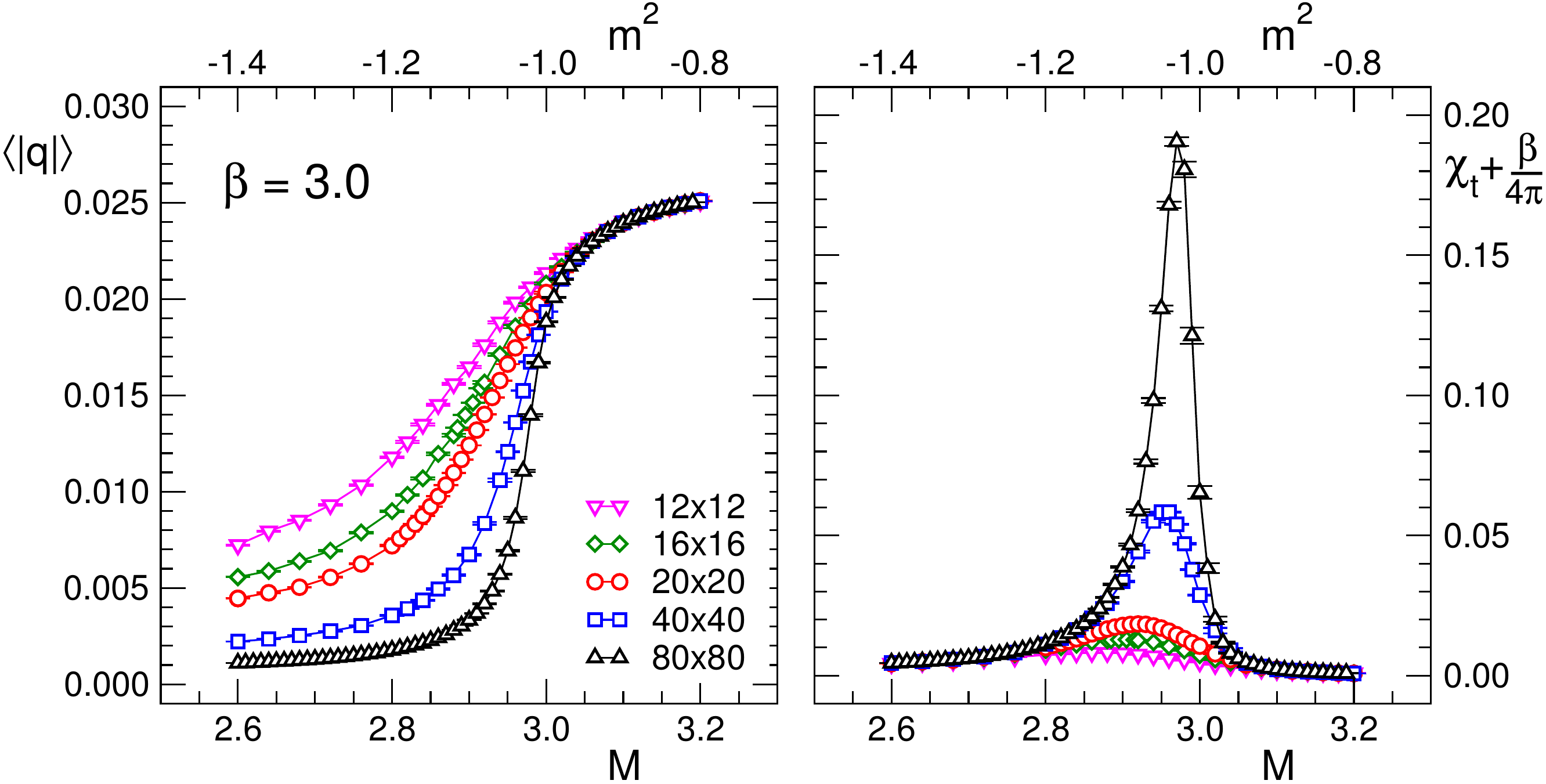}
\vskip5mm
\includegraphics[width=125mm,clip]{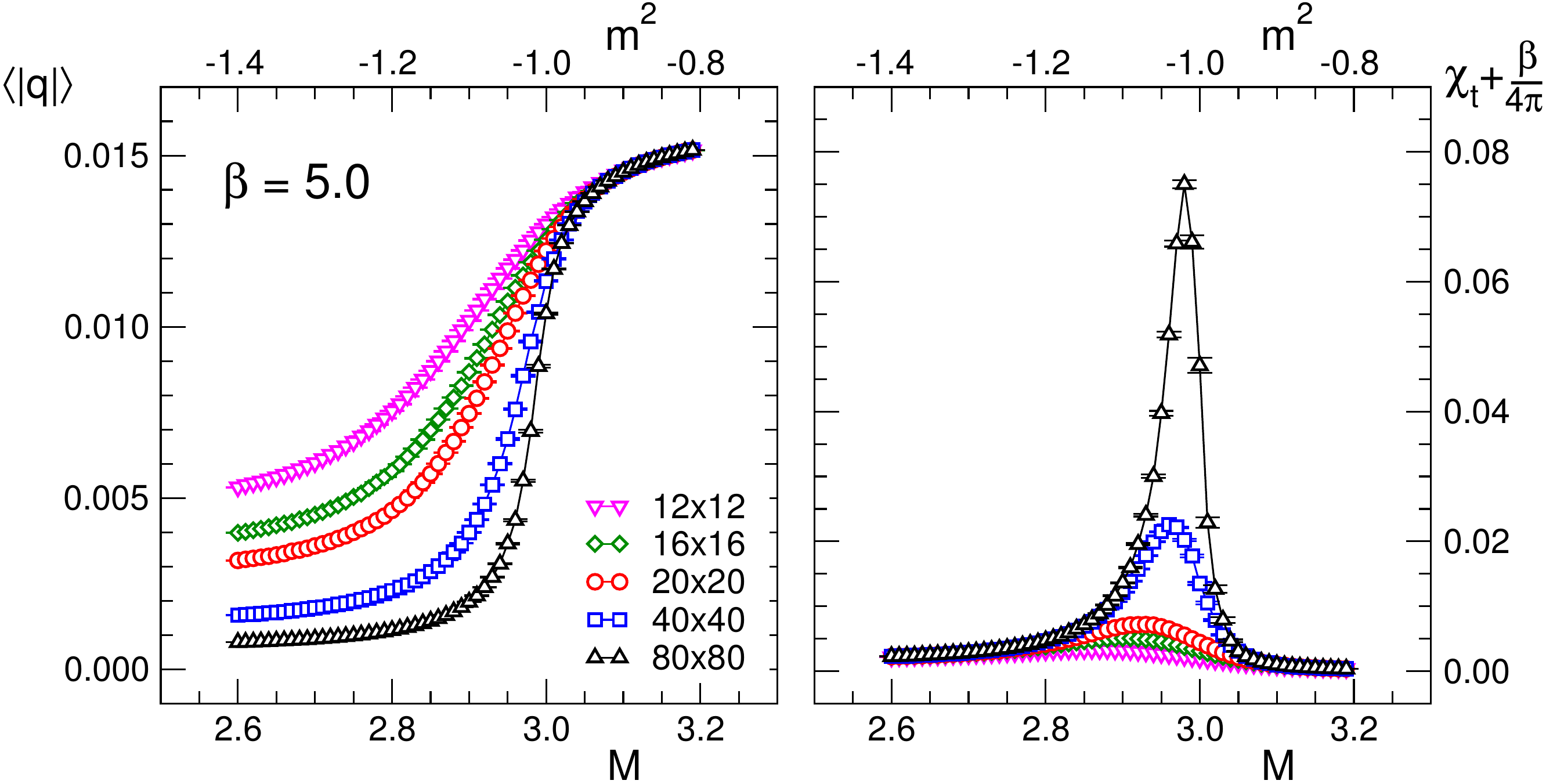}
\caption{The topological charge density $\langle | q | \rangle$ and the topological susceptibility shifted by a 
constant $\chi_t+\frac{\beta}{4\pi}$ at $\theta = \pi$ for different volumes. We compare the results for 
$\beta = 3.0$ (top plots) and $\beta = 5.0$ (bottom) at $\lambda = 0.5$ 
and plot the observables as a function of our control parameter, i.e., the mass parameter $M = 4 + m^2$. 
For comparison on the top of the plots we also show the horizontal axis labelled with $m^2$, 
which at the point of the transition is negative.}
\label{q_chi_vs_M}
\end{figure}

In order to locate and analyze the phase transition we study $\langle | q | \rangle$ and $\chi_t$ as a function of 
the control parameter $M = 4 + m^2$. In Fig.~\ref{q_chi_vs_M} we show the corresponding results for $\lambda = 0.5$
and $\beta = 3.0$ (top plots), as well as $\beta = 5.0$, and compare the data for different lattice sizes. 
As already anticipated from our comparison 
of $q$ to the magnetization in a ferromagnet, we indeed observe the characteristic behavior expected for a symmetry 
breaking transition. At a critical mass parameter of $M_c \sim 3.0$ we observe the onset of the expectation value  
$\langle | q | \rangle$, which is of course rounded by finite volume effects, and only for infinite volume approaches
the expected form of $\langle | q | \rangle = 0$ for $M < M_c$, a non-differentiable edge at $M_c$ and a quick 
approach to a constant behavior above $M_c$. Comparing the plots for $\beta = 3.0$ and $\beta = 5.0$ we note 
that the value for the critical mass parameter $M_c$ shows only a weak dependence on the inverse 
gauge coupling $\beta$. 

Also the corresponding susceptibility $\chi_t$ shown in the rhs.\ plot of Fig.~\ref{q_chi_vs_M} displays the expected 
behavior of a growing maximum as the volume is increased. The rate at which this maximum grows with the extent 
$L$ of the lattice ($N_S = N_T = L$ here) is expected to be $\chi_{t,max} \sim L^{\gamma/\nu} = L^{7/4}$ 
for the conjectured Ising transition. Despite the fact that for a precise determination of the maxima a much finer 
spacing of the data points would be needed, it is already plausible that the maxima indeed scale as conjectured. 
In the next section we now determine the critical mass parameter $M_c$ as well as some of the critical exponents
using precise finite site scaling techniques.

\subsection{Definition of the critical exponents and the strategy for their determination from finite size scaling}

In this subsection we define the critical exponents we want to determine and briefly summarize the 
setting of our finite scaling analysis. This is of course text book material (see, e.g., 
\cite{Newman_Barkema,Landau_Binder}), but needs to be slightly adapted to the observables of the 
U(1) gauge Higgs model studied here. 

In our system the mass parameter $M$ is the quantity that drives the system through the phase transition at the 
critical value $M_c$. Consequently we define the reduced mass $m_r$,
\begin{equation}
\label{reducedm}
m_r \; \equiv \;  \frac{M-M_c}{M_c} \; .
\end{equation}
We already discussed that the topological charge density $q$ has to be identified with the magnetization, and the 
topological susceptibility $\chi_t$ with the magnetic susceptibility. Near the transition at $m = 0$ 
these observables behave as
\begin{equation}
\label{critical_exps}
\langle | q | \rangle_{(m_r)} \;  \sim \; |m_r|^{\beta} \; \; , \qquad
\chi_{t \, (m_r)} \; \sim \; |m_r|^{-\gamma} \; \; , \qquad  \xi_{(m_r)} \; \sim \;  |m_r|^{-\nu} \; ,
\end{equation} 
where $\xi$ is the correlation length. The terminology for the corresponding critical exponents $\beta$, $\gamma$ and
$\nu$ follows the conventions for ferromagnets. Using the third equation in (\ref{critical_exps}) to express $|m_r|$ in terms
of $\xi$ and the fact that the correlation length is limited by $L$ (we use lattices with size $N_S = N_T = L$), we can scale
out the leading term in  $\langle | q | \rangle$ and $\chi_t$ to find the finite size scaling form
\begin{equation}
\langle | q | \rangle_{(m_r,L)} \; = \; L^{-\frac{\beta}{\nu}} \ F_{q} (x) \; \; , \qquad
\chi_{t \, (m_r,L)} \; = \; L^{\frac{\gamma}{\nu}} \ F_{\chi} (x) \; \; ,
\label{FSS1}
\end{equation}
where $x = L^{\frac{1}{\nu}} \, m_r$ is the scaling variable and $F_{q} (x)$ and $F_{\chi} (x)$ are 
the scaling functions that are regular at $m_r = 0$ and have no further $L$ dependence in the scaling region 
around $m_r =0$. 

For the determination of the critical exponents we implement the Ferren-berg-Landau strategy 
\cite{Ferrenberg_Landau:1991}. We introduce the fourth order Binder cumulant \cite{Binder1981}
\begin{equation}
\label{equ:binder_cumulant}
U \; \equiv \;  1 \, - \, \frac{\langle \, q^{\,4} \, \rangle}{\, 3 \; \langle \, q^{\,2} \, \rangle^2} \; .
\end{equation}
The maximum slope of this cumulant scales with $L^\frac{1}{\nu}$, such that the peak of its mass-derivative scales as \cite{Binder1981}
\begin{equation}
\label{equ:binder_scaling}
\frac{dU}{dM}\Big|_\text{max} \; = \; A \, L^\frac{1}{\nu} \; .
\end{equation}
The first step of our determination of the critical exponents (see the next subsection for the 
discussion of the actual implementation) is the determination of $\nu$ based on 
(\ref{equ:binder_scaling}), with a numerically computed derivative $dU / dM$ of the Binder cumulant based on a 
multiple-histogram technique (see below). 

In addition we also determine $\nu$ by studying observables that have 
the same scaling behavior as $U$, for instance the logarithmic derivatives of  moments of the topological charge. 
In particular we study the derivatives 
\begin{equation}
	\frac{d}{dM} \ln \, \langle |q|\rangle \; = \; \frac{1}{\langle |q|\rangle} \frac{d}{dM} \langle |q|\rangle \; , \qquad
	\frac{d}{dM} \ln\, \langle q^2\rangle \; = \; \frac{1}{\langle q^2\rangle} \frac{d}{dM} \langle q^2\rangle \; , 
	\label{equ:diff_ln_obs}
\end{equation}
which have maxima that also scale with $L^\frac{1}{\nu}$. Thus these derivatives can again be fit as
described in (\ref{equ:binder_scaling}) and allow for an independent determination of $\nu$.

Once the exponent $\nu$ has been calculated with satisfactory precision we can determine the critical mass $M_c$. 
We first estimate the pseudo-critical mass values $M_{pc}(L)$ by determining the location of the peaks in observables 
evaluated at lattice size $L$. Suitable observables for this analysis are the derivative $dU/dM$ 
(\ref{equ:binder_scaling}), the logarithmic derivatives (\ref{equ:diff_ln_obs}) as well as the location of the maximum 
slope in $\langle|q|\rangle$ and the peak in the susceptibility $\chi_{t}$. The critical mass $M_c$ was then  
determined in a fit of the data for the pseudo-critical mass values $M_{pc}(L)$ with the scaling ansatz
\begin{equation}
	\label{equ:estimate_Mc}
	M_{pc}(L) \; = \; M_c \; + \; A\, L^{-\frac{1}{\nu}}  \; .
\end{equation}

Once $\nu$ and $M_c$ are determined, we know the zero of the reduced mass $m$ defined in (\ref{reducedm}) 
and the scaling variable $x = L^{\frac{1}{\nu}} \, m_r$. Thus we can compute the remaining critical exponents 
from the finite size scaling relations (\ref{FSS1}). $\beta$ is determined from the scaling of  $\langle|q|\rangle_{(m_r,L)}$,
and $\gamma$ from the scaling of $\chi_{t \, (m_r,L)}$.

To reliable determine the precise locations $M_{pc}(L)$ of the peaks in observables we use the multiple-histogram 
method \cite{Multihisto}. This approach allows one to perform only a few simulations 
in the region around the peaks and to compute the observables at a continuum of values $M$ with high precision. 
An error estimate for the necessary interpolation can be obtained by jackknife resampling of the data.

\subsection{Numerical results for the critical exponents}

For the determination of the critical exponents with the finite size scaling strategy outlined in the previous subsection 
we use $L \times L$ lattices with sizes of $L =$ 16, 32, 48, 64, 80, 96 and 112 and an inverse gauge coupling of 
$\beta = 3.0$. As already outlined, for each lattice size $L$ we perform simulations at 4 different values of the 
mass $M$ within the critical region and interpolate with fine interpolation steps using the multiple histogram method 
\cite{Multihisto}. For an error estimation we divide our measurements into 
blocks and perform a jackknife resampling of the data blocks. The critical exponents and the critical mass are then 
extracted following the strategy outlined in the previous subsection. For the simulations to determine the critical 
exponents we use ensembles with $4\times 10^6$ configurations after $5\times 10^4$ initial equilibration sweeps and 
$50$ decorrelation sweeps between the measurements.

\begin{figure}[t!]
	\centering
	\includegraphics[width=60.3mm,clip]{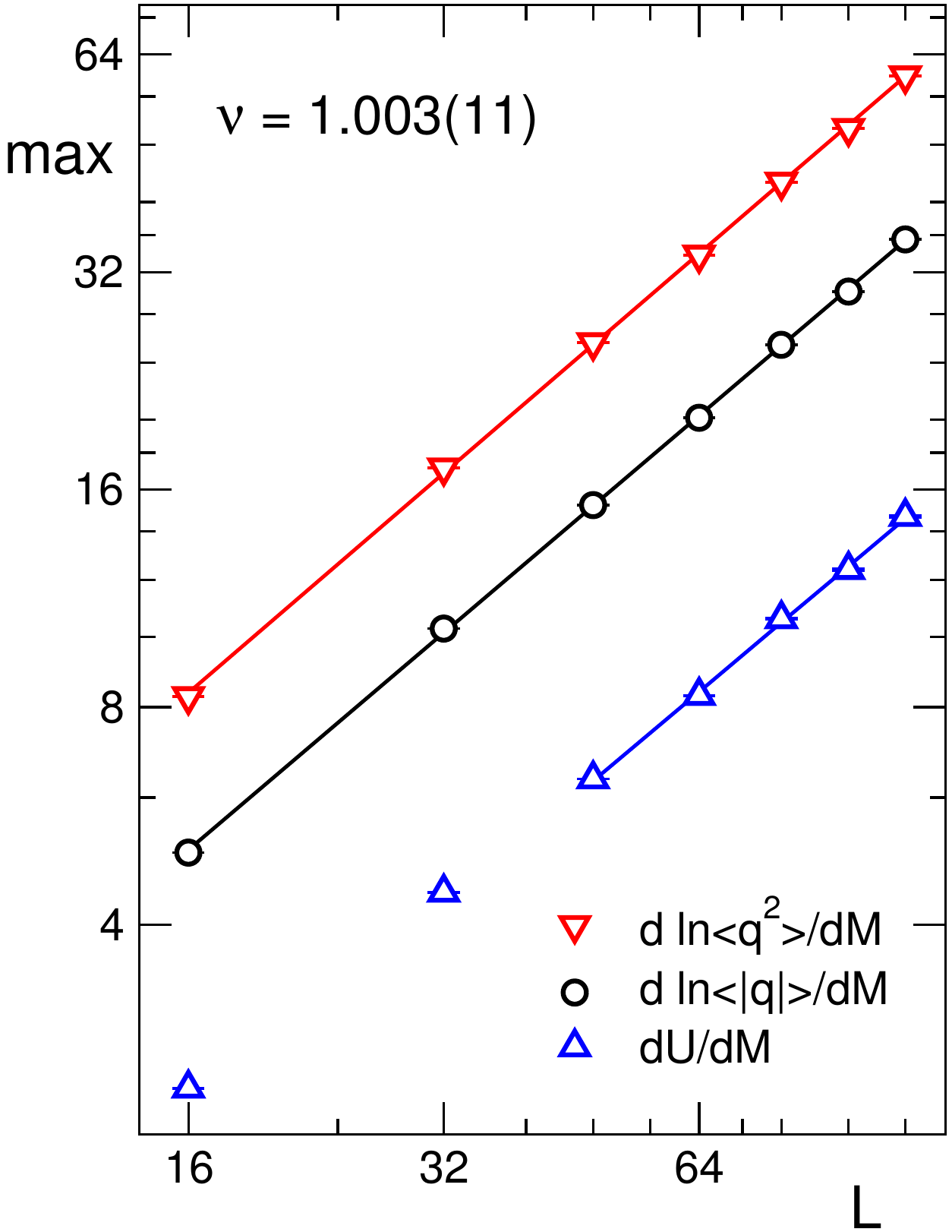}
	\includegraphics[width=65mm,clip]{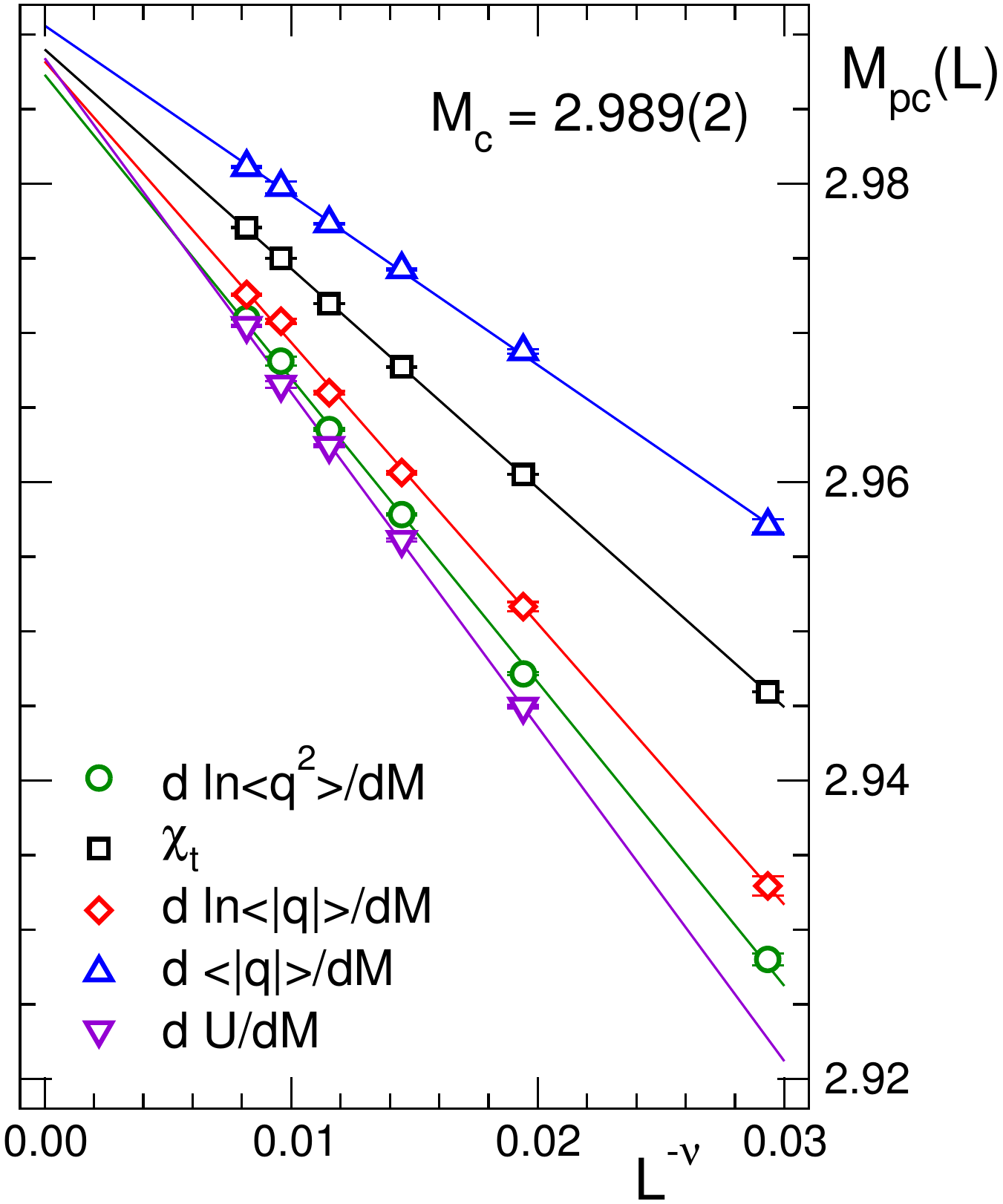}
	\caption{Lhs.: The maxima of the mass-derivative of the Binder cumulant $U$ and the logarithmic 
	derivatives of $\langle |q| \rangle$ and $\langle q^2 \rangle$ plotted as a function of the lattice size $L$. 
	The solid lines show fits with $A \, L^{\frac{1}{\nu}}$ which were used to determine $\nu$.
	Rhs.: The pseudo-critical values $M_{pc}(L)$ determined as the positions of the maxima of different observables
	(see the labels in the legend), plotted against $L^{-\nu}$. The solid lines are fits according to 
	(\ref{equ:estimate_Mc}) and their values extrapolated to $L = \infty$ were used for the determination of $M_c$. In 
	both plots we annotate the final results for the parameters determined from the fits.}
	\label{plot_nu_Mc}
\end{figure}

The first step in the analysis is the determination of the critical exponent $\nu$, which can be performed independent of 
other critical exponents and without the knowledge of the critical mass $M_c$. As discussed we determine 
$\nu$ from the maxima of the mass-derivatives of $U$, $\ln \, \langle |q|\rangle$ and $\ln \, \langle q^2 \rangle$ 
(compare Eqs.~(\ref{equ:binder_scaling}) and (\ref{equ:diff_ln_obs})). In the lhs.~plot of Fig.~\ref{plot_nu_Mc} the symbols 
indicate the maxima of the mass derivatives of the three observables plotted against the lattice size. 
In our double logarithmic plot the maxima fall onto straight lines that are essentially parallel and have a slope 
given by $1/\nu$. From a fit of the maxima with the scaling behavior (\ref{equ:binder_scaling}) we determine $\nu$.

\begin{figure}[t!]
	\centering
	\includegraphics[width=62mm,clip]{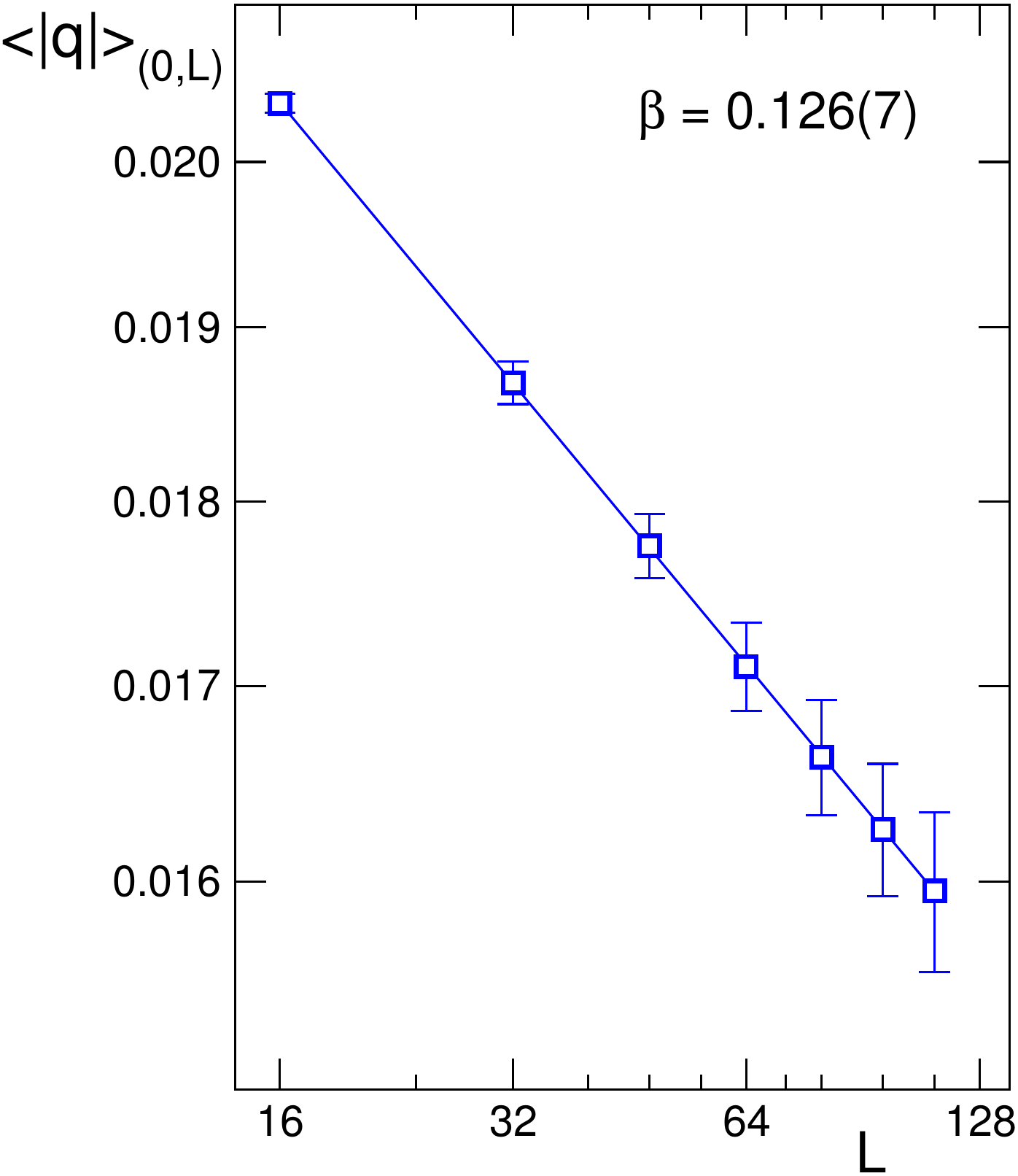}
	\includegraphics[width=59.6mm,clip]{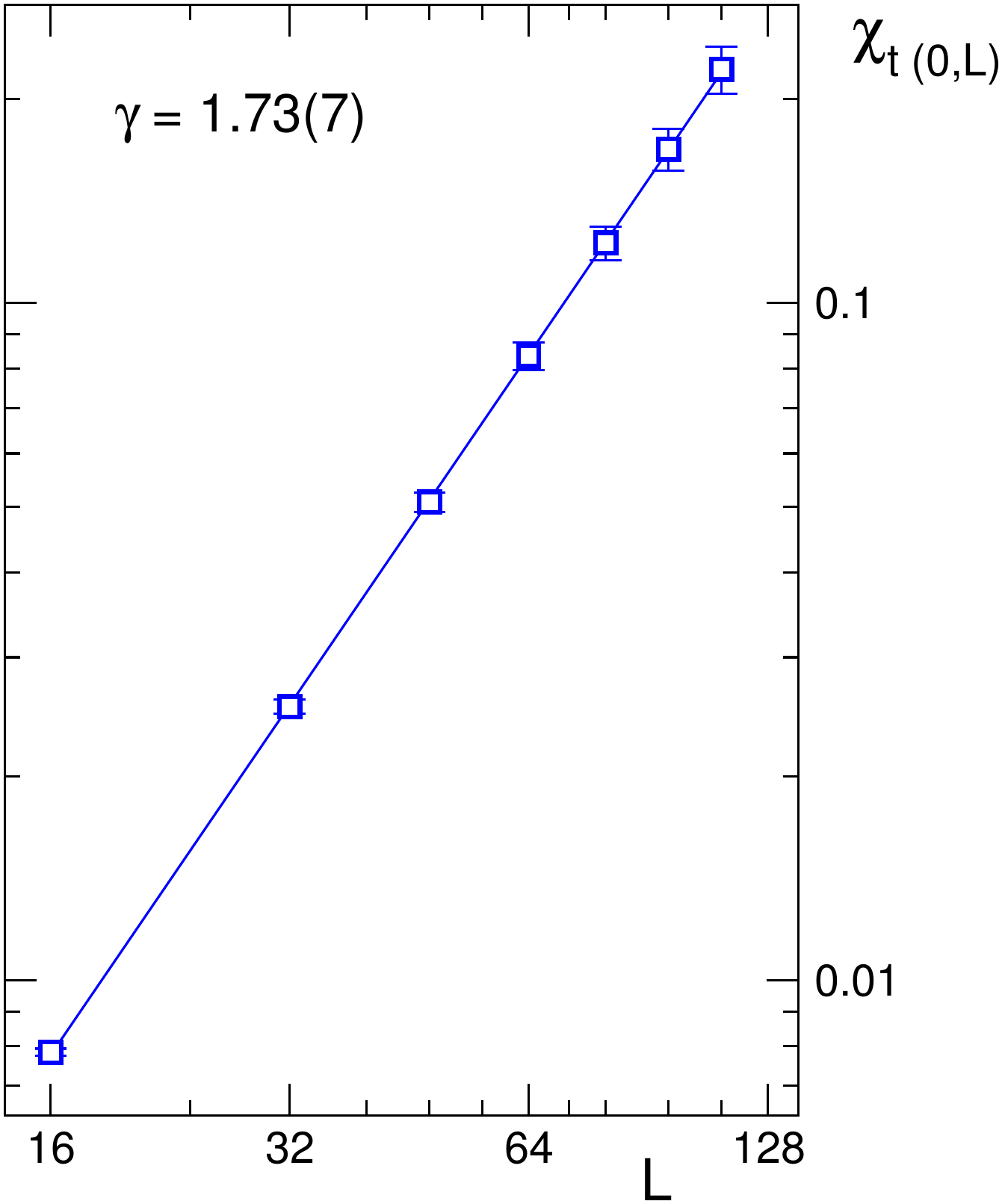}
	\caption{Lhs.: Scaling of the topological charge $\langle |q| \rangle$. The symbols indicate the values of 
	$\langle |q| \rangle_{(0,L)}$ at the critical mass $M_c$ (i.e., at reduced mass $m_r = 0$) for different lattice sizes. 
	The solid line is the fit to these data according to (\ref{FSS1}). 
	Rhs.: Scaling of the topological susceptibility $\chi_t$. The symbols show the values of $\chi_{t\,(0,L)}$ at the 
	critical mass for different lattice sizes. The solid line is again the fit to the finite size scaling formula (\ref{FSS1}). In 
	both plots we annotate the final results for the parameters determined from the fits.}
	\label{plot_beta_gamma}
\end{figure}

By performing a series of fits where we vary the minimal size $L$ taken into account in the fit range we can see that 
up to $L=32$ the Binder cumulant is sensitive to additional finite size effects and we thus omit the corresponding data 
for the smallest two lattice sizes from the fit, while for the other observables all measurements could be used. The 
values of $\nu$ extracted from the fits of the different observables were combined in an average leading to a 
final result of $\nu = 1.003(11)$. The statistical error was propagated through the entire analysis chain 
starting with the initial jackknife resampling of the data.  

With the knowledge of the critical exponent $\nu$ we can now determine the critical mass $M_c$ from the scaling 
relation (\ref{equ:estimate_Mc}) for the pseudo-critical values $M_{pc}(L)$ obtained as the maxima of suitable 
observables. As outlined, we use the mass derivatives of $U$, $\langle|q|\rangle$, $\ln \langle|q|\rangle$ and 
$\ln \langle q^2\rangle$ as well as the location of the peaks in $\chi_t$. The symbols in the rhs.\ plot of 
Fig.~\ref{plot_nu_Mc} show the corresponding values $M_{pc}(L)$ as a function of $L^{-\nu}$, and the straight lines 
the fits with (\ref{equ:estimate_Mc}). We indeed observe the expected linear behavior and the fits for the different 
observables extrapolate for $L \rightarrow \infty$ to the corresponding estimates for $M_c$. Again we varied the fit 
range to analyze higher order finite size corrections and discarded the $L=16$ data for all observables and in 
addition removed the $L=32$ data for the Binder cumulant. The extrapolated results for the critical mass were 
then averaged with a final result of $M_c = 2.989(2)$. 
 
After estimating $\nu$ and $M_c$ we can finally extract the exponents $\beta$ and $\gamma$ from the scaling of 
$\langle|q|\rangle_{(m_r,L)}$ and $\chi_{t \, (m_r,L)}$, as given in (\ref{FSS1}). In the double logarithmic plots in 
Fig.~\ref{plot_beta_gamma} we show the results for $\langle|q|\rangle_{(m_r,L)}$ and $\chi_{t \, (m_r,L)}$ at $M_c$, i.e., 
at $m_r = 0$ as a function of $L$. Fits according to the scaling relations (\ref{FSS1}) appear as straight lines and their 
slopes are given by $- \beta/\nu$ and by $\gamma/\nu$. Using the previously determined result for $\nu$ we obtain 
values of $\beta = 0.126(7)$ and $\gamma = 1.73(7)$. 

This concludes our determination of the critical exponents and the critical mass parameter $M$. 
We summarize our results in Table~\ref{tab:table1}, where we display also the corresponding critical 
values of the 2d Ising model. We observe good agreement within error bars, which provides strong numerical evidence 
that the critical endpoint of the 2d gauge Higgs model at $\theta = \pi$ is in the 2d Ising universality class. 
\vskip5mm

\begin{table}[h!]
	\begin{center}
		\begin{tabular}{ccc} 
			\hline
			& & \vspace{-2.5mm}\\
			\textbf{Parameter} \quad & \quad \textbf{Numerical result} \quad & \quad \textbf{2d Ising} \vspace{1.5mm} \\
			\hline 
			& & \vspace{-2.5mm}\\
			$M_C$ & 2.989(2) &  \\
			$\nu$ & 1.003(11) & 1 \\
			$\beta$ & 0.126(7) &  0.125 \\
			$\gamma$ & 1.73(7) & 1.75 \\
			& & \vspace{-2.5mm}\\
			\hline 
		\end{tabular}
	\end{center}
	\caption{Numerical results for the critical mass $M_C$ and the critical exponents 
	$\nu$, $\beta$ and $\gamma$. In the third column we list the critical exponents of the two dimensional 
	Ising model for comparison.}
	\label{tab:table1}
\end{table}

\section{Summary and future prospects}

In this paper we have presented the results of our lattice analysis of the critical endpoint for the 2d U(1) 
gauge Higgs model at topological angle $\theta = \pi$, where charge conjugation is an exact symmetry. Charge 
conjugation is spontaneously broken when the scalars with a gauge charge are heavy. As the scalar mass squared 
is lowered and becomes sufficiently negative, the symmetry can become unbroken with a transition that is conjectured 
to be in the 2d Ising universality class. 

For studying this scenario numerically, the lattice formulation has to deal with two main challenges, the implementation 
of the topological charge as an integer in order to make charge conjugation an exact symmetry at $\theta = \pi$, and 
the complex action problem at non-zero topological angle. The former problem is overcome here by using the Villain
action, while the latter challenge is solved by using a dual representation in terms of worldlines 
and plaquette occupation numbers for the simulation. 

Using the topological charge density $q$, and the corresponding topological susceptibility $\chi_t$ as our main 
observables we show that for sufficiently large mass parameter $M = 4 + m^2$ the system undergoes a first order 
transition at $\theta = \pi$. As the mass parameter is lowered this first order behavior disappears in a critical 
endpoint $M_c$. As a function of $M$ the observables $q$ and $\chi_t$ indicate a second order transition at $M_c$. 
Using finite size scaling analysis we determine the corresponding critical exponents and show that the transition is in 
the 2d Ising universality class as conjectured.

The work presented here is only the first step in exploring the rich physics of 2d quantum field theories at finite topological 
$\theta$-angle. The methods presented here can be adopted for studying theories with more scalar matter fields. 
If the fields have the same mass, the global symmetry group is enhanced, and there exist nontrivial 't~Hooft 
anomaly matching conditions at $\theta=\pi$, which guarantee that there is no trivially gapped phase (see 
\cite{Komargodski:2017dmc,Komargodski:2017smk,Tanizaki:2017qhf,Tanizaki:2018xto,Sulejmanpasic:2018upi}). Another 
interesting case is to promote the charge-1 scalar field to a charge-2 scalar field. Such a theory, in addition to the 
$C$-symmetry has a $\mathds Z_2$ center symmetry, which leads to a mixed 't Hooft anomaly. 
This model, which can be studied on the lattice with our methods, will exhibit a 
very different behavior\footnote{The behavior at large and positive/negative $m^2$ can be computed semiclassically and and it shows that at both ends $C$-symmetry is spontaneously broken. See, e.g., the discussion in 
\cite{Sulejmanpasic:2018upi}.} from the charge-1 model. 

Similar reasoning can also be used to formulate compact $U(1)$ gauge theories in higher dimensions. By doing this, the 
theory of compact electrodynamics can be formulated which allows for an exclusion of dynamical monopoles as well as 
complete control over their contributions to a partition sum (charges, actions, etc.).  One application of such a formulation 
is that a theory with only fixed monopole charges (typically multiples of $2$ or $4$) can be formulated, or indeed even 
without any monopoles. Such theories are the effective theories of anti-ferromagnets, and have been connected with 
deconfined quantum criticality\footnote{The connection with quantum magnets is only for theories with multiple scalar 
flavors, similar to the case of 2d.} and emergent symmetries at the phase transition between the monopole phase, and 
the Higgs phase. 

Finally our approach can also be adopted for studying various non-linear sigma models, such as $SU(N)/U(1)^{N-1}$, 
which have $N-1$ $\theta$-angles and can be formulated in terms of $U(1)^{N-1}$ scalar gauge theories. For $N=2$ 
this is the famous $O(3)$ or $CP(1)$ nonlinear sigma model, which at $\theta=\pi$ is an effective theory of half-integer 
spin chains, while for $N=3$ the model corresponds to an effective theory of $SU(3)$ spin chains 
\cite{Bykov:2011ai,Bykov:2012am,Lajko:2017wif}. Being asymptotically free, the model also has two 
$\theta$-angles, and a rich phase 
diagram as a function of the two angles (see \cite{Lajko:2017wif,Tanizaki:2018xto}) which can be explored with the 
methods presented here\footnote{We note that the application of the methods presented here is not completely 
straightforward because of the absence of the gauge-kinetic terms in the formulations of the non-linear sigma models via 
gauge fields. This is undesirable as it allows lattice-scale instantons to overwhelm the physics of the $\theta$-angle. This 
problem can however be avoided by adding a gauge-kinetic term to the action to ensure smoothness of gauge fields on 
the lattice scale. Keeping its coupling fixed, the term vanishes in the continuum limit.}.

\vskip5mm
\noindent
{\bf Acknowledgments:}
\vskip2mm
\noindent
We would like to thank Yuya Tanizaki for his comments on the draft. 
This work is supported by the Austrian Science Fund FWF, grant I 2886-N27 and the FWF
DK ''Hadrons in Vacuum, Nuclei and Stars'', grant W-1203. 
Parts of the computational results presented have been achieved using the Vienna Scientific Cluster (VSC).
The authors would like to express a special thanks to the Mainz Institute for Theoretical Physics (MITP) for its hospitality 
and support during the workshop ''Progress in Diagrammatic Monte Carlo Methods'' which facilitated this collaboration.

\vskip5mm
\section*{Appendix}

In order to keep this paper self-contained, in this appendix we derive the Fourier representation of the Boltzmann 
factor $B_G[U]$ which we use in (\ref{B_G_dual}). In its conventional form the Boltzmann factor 
with Villain action is the product over factors
\begin{equation}
b(F) \; = \; \sum_{n \in \mathds{Z}} 
e^{\, - \, \frac{\beta}{2}\left(F+2\pi n\right)^2 \, - \, i \frac{\theta}{2\pi}\left(F+2\pi n\right)} \; ,
\end{equation}
at each plaquette labelled by its lower left corner $x$ (we here omit the corresponding index $x$ for brevity).
Obviously the function $b(F)$ is $2\pi$-periodic in $F$ such that it has the Fourier representation 
\begin{equation}
b(F) \; = \; \sum_{p \in \mathds{Z}} \, \widetilde{b}(p) \;\, e^{i p F} \; .
\label{Frep}
\end{equation}
The Fourier transform $\widetilde{b}(p)$ can be evaluated in closed form,
\begin{eqnarray}
\widetilde{b}(p) & \!=\! & \int_{-\pi}^\pi \!\!\! dF \; b(F) \; e^{\,-i p F} \; = \; 
\sum_{n \in \mathds{Z}} \, \int_{-\pi}^\pi \!\!\! dF \; 
e^{\, - \, \frac{\beta}{2}\left(F+2\pi n\right)^2 \, - \, i \frac{\theta}{2\pi}\left(F+2\pi n\right)} \; e^{\,-i p F} 
\nonumber \\
& \!=\! & 
\sum_{n \in \mathds{Z}} \, \int_{-\pi}^\pi \!\!\! dF \; 
e^{\, - \, \frac{\beta}{2}\left(F+2\pi n\right)^2 \, - \, i \frac{\theta}{2\pi}\left(F+2\pi n\right) \,- \, i p \left(F+2\pi n\right)} 
\; = \; \Big| F^\prime \!= F\!+\!2\pi n \Big|
\nonumber \\
& \!=\! & 
\sum_{n \in \mathds{Z}} \, \int_{2\pi n -\pi}^{2\pi n + \pi} \!\! dF^\prime \; 
e^{\, - \, \frac{\beta}{2}\left(F^\prime\right)^2 \, - \, i F^\prime\left(p + \frac{\theta}{2\pi}\right)} 
\nonumber \\
& \!=\! & 
\int_{-\infty}^{\infty} \!\! dF^\prime \; 
e^{\, - \, \frac{\beta}{2}\left(F^\prime\right)^2 \, - \, i F^\prime\left(p + \frac{\theta}{2\pi}\right)}
\; = \; \frac{1}{\sqrt{2\pi \beta}} \; e^{- \frac{1}{2\beta} \left(p + \frac{\theta}{2\pi}\right)^2} \; .
\end{eqnarray}
Inserting this Fourier transform $\widetilde{b}(p)$ in (\ref{Frep}) gives rise to the dual representation 
(\ref{B_G_dual}) of the Boltzmann factor $B_G[U]$.

\vskip10mm
\bibliographystyle{utphys}
\bibliography{bibliography}

\providecommand{\href}[2]{#2}\begingroup\raggedright\begin{thebibliography}{10}

\bibitem{Gaiotto:2017yup}
D.~Gaiotto, A.~Kapustin, Z.~Komargodski, and N.~Seiberg, ``{Theta, Time
  Reversal, and Temperature},'' {\em JHEP} {\bf 05} (2017) 091,
  \href{http://xxx.lanl.gov/abs/1703.00501}{{\tt 1703.00501}}.

\bibitem{Gaiotto:2017tne}
D.~Gaiotto, Z.~Komargodski, and N.~Seiberg, ``{Time-reversal breaking in
  QCD$_{4}$, walls, and dualities in 2 + 1 dimensions},'' {\em JHEP} {\bf 01}
  (2018) 110, \href{http://xxx.lanl.gov/abs/1708.06806}{{\tt 1708.06806}}.

\bibitem{DiVecchia:2017xpu}
P.~Di~Vecchia, G.~Rossi, G.~Veneziano, and S.~Yankielowicz, ``{Spontaneous $CP$
  breaking in QCD and the axion potential: an effective Lagrangian approach},''
  {\em JHEP} {\bf 12} (2017) 104,
  \href{http://xxx.lanl.gov/abs/1709.00731}{{\tt 1709.00731}}.

\bibitem{Komargodski:2017dmc}
Z.~Komargodski, A.~Sharon, R.~Thorngren, and X.~Zhou, ``{Comments on Abelian
  Higgs Models and Persistent Order},''
  \href{http://xxx.lanl.gov/abs/1705.04786}{{\tt 1705.04786}}.

\bibitem{Komargodski:2017smk}
Z.~Komargodski, T.~Sulejmanpasic, and M.~{\"U}nsal, ``{Walls, anomalies, and
  deconfinement in quantum antiferromagnets},'' {\em Phys. Rev.} {\bf B97}
  (2018), no.~5 054418, \href{http://xxx.lanl.gov/abs/1706.05731}{{\tt
  1706.05731}}.

\bibitem{Tanizaki:2017qhf}
Y.~Tanizaki, T.~Misumi, and N.~Sakai, ``{Circle compactification and 't~Hooft
  anomaly},'' {\em JHEP} {\bf 12} (2017) 056,
  \href{http://xxx.lanl.gov/abs/1710.08923}{{\tt 1710.08923}}.

\bibitem{Tanizaki:2018xto}
Y.~Tanizaki and T.~Sulejmanpasic, ``{Anomaly and global inconsistency matching:
  $\theta$-angles, $SU(3)/U(1)^2$ nonlinear sigma model, $SU(3)$ chains and its
  generalizations},'' \href{http://xxx.lanl.gov/abs/1805.11423}{{\tt
  1805.11423}}.

\bibitem{Sulejmanpasic:2018upi}
T.~Sulejmanpasic and Y.~Tanizaki, ``{C-P-T anomaly matching in bosonic quantum
  field theory and spin chains},'' {\em Phys. Rev.} {\bf B97} (2018), no.~14
  144201, \href{http://xxx.lanl.gov/abs/1802.02153}{{\tt 1802.02153}}.

\bibitem{Endres:2006xu}
M.~G. Endres, ``{Method for simulating O(N) lattice models at finite
  density},'' {\em Phys. Rev.} {\bf D75} (2007) 065012,
  \href{http://xxx.lanl.gov/abs/hep-lat/0610029}{{\tt hep-lat/0610029}}.

\bibitem{Chandrasekharan:2008gp}
S.~Chandrasekharan, ``{A New computational approach to lattice quantum field
  theories},'' {\em PoS} {\bf LATTICE2008} (2008) 003,
  \href{http://xxx.lanl.gov/abs/0810.2419}{{\tt 0810.2419}}.

\bibitem{Mercado:2013yta}
Y.~Delgado~Mercado, C.~Gattringer, and A.~Schmidt, ``{Surface worm algorithm
  for abelian Gauge-Higgs systems on the lattice},'' {\em Comput. Phys.
  Commun.} {\bf 184} (2013) 1535--1546,
  \href{http://xxx.lanl.gov/abs/1211.3436}{{\tt 1211.3436}}.

\bibitem{Gattringer:2012jt}
C.~Gattringer and A.~Schmidt, ``{Gauge and matter fields as surfaces and loops
  - an exploratory lattice study of the Z(3) Gauge-Higgs model},'' {\em Phys.
  Rev.} {\bf D86} (2012) 094506, \href{http://xxx.lanl.gov/abs/1208.6472}{{\tt
  1208.6472}}.

\bibitem{Korzec:2012fa}
T.~Korzec and U.~Wolff, ``{Simulating the All-Order Strong Coupling Expansion
  V: Ising Gauge Theory},'' {\em Nucl. Phys.} {\bf B871} (2013) 145--163,
  \href{http://xxx.lanl.gov/abs/1212.2875}{{\tt 1212.2875}}.

\bibitem{Mercado:2013ola}
Y.~Delgado~Mercado, C.~Gattringer, and A.~Schmidt, ``{Dual Lattice Simulation
  of the Abelian Gauge-Higgs Model at Finite Density: An Exploratory Proof of
  Concept Study},'' {\em Phys. Rev. Lett.} {\bf 111} (2013), no.~14 141601,
  \href{http://xxx.lanl.gov/abs/1307.6120}{{\tt 1307.6120}}.

\bibitem{Korzec:2013dra}
T.~Korzec and U.~Wolff, ``{Simulating the Random Surface representation of
  Abelian Gauge Theories},'' {\em PoS} {\bf LATTICE2013} (2014) 038,
  \href{http://xxx.lanl.gov/abs/1311.5198}{{\tt 1311.5198}}.

\bibitem{Kloiber:2014dfa}
T.~Kloiber and C.~Gattringer, ``{Scalar QED$_2$ with a topological term -- a
  lattice study in a dual representation},'' {\em PoS} {\bf LATTICE2014} (2014)
  345.

\bibitem{Schmidt:2015cva}
A.~Schmidt, P.~de~Forcrand, and C.~Gattringer, ``{Condensation phenomena in
  two-flavor scalar QED at finite chemical potential},'' {\em PoS} {\bf
  LATTICE2014} (2015) 209, \href{http://xxx.lanl.gov/abs/1501.06472}{{\tt
  1501.06472}}.

\bibitem{Gattringer:2015baa}
C.~Gattringer, T.~Kloiber, and M.~M{\"u}ller-Preussker, ``{Dual simulation of
  the two-dimensional lattice U(1) gauge-Higgs model with a topological
  term},'' {\em Phys. Rev.} {\bf D92} (2015), no.~11 114508,
  \href{http://xxx.lanl.gov/abs/1508.00681}{{\tt 1508.00681}}.

\bibitem{Bruckmann:2015sua}
F.~Bruckmann, C.~Gattringer, T.~Kloiber, and T.~Sulejmanpasic, ``{Dual lattice
  representations for O(N) and CP(N-1) models with a chemical potential},''
  {\em Phys. Lett.} {\bf B749} (2015) 495--501,
  \href{http://xxx.lanl.gov/abs/1507.04253}{{\tt 1507.04253}}. [Erratum: Phys.
  Lett.B751,595(2015)].

\bibitem{Bruckmann:2015hua}
F.~Bruckmann, C.~Gattringer, T.~Kloiber, and T.~Sulejmanpasic, ``{Grand
  Canonical Ensembles, Multiparticle Wave Functions, Scattering Data, and
  Lattice Field Theories},'' {\em Phys. Rev. Lett.} {\bf 115} (2015), no.~23
  231601, \href{http://xxx.lanl.gov/abs/1509.05189}{{\tt 1509.05189}}.

\bibitem{Villain:1974ir}
J.~Villain, ``{Theory of one-dimensional and two-dimensional magnets with an
  easy magnetization plane. 2. The Planar, classical, two-dimensional
  magnet},'' {\em J. Phys.(France)} {\bf 36} (1975) 581--590.

\bibitem{Gaiotto:2014kfa}
D.~Gaiotto, A.~Kapustin, N.~Seiberg, and B.~Willett, ``{Generalized Global
  Symmetries},'' {\em JHEP} {\bf 02} (2015) 172,
  \href{http://xxx.lanl.gov/abs/1412.5148}{{\tt 1412.5148}}.

\bibitem{Prokofev:2001ddj}
N.~Prokof'ev and B.~Svistunov, ``{Worm Algorithms for Classical Statistical
  Models},'' {\em Phys. Rev. Lett.} {\bf 87} (2001) 160601.

\bibitem{Wiese:1988qz}
U.~J. Wiese, ``{Numerical Simulation of Lattice $\theta$ Vacua: The 2-$d$ U(1)
  Gauge Theory as a Test Case},'' {\em Nucl. Phys.} {\bf B318} (1989) 153--175.

\bibitem{Newman_Barkema}
M.~Newman and G.~Barkema, {\em Monte Carlo Methods in Statistical Physics}.
\newblock Clarendon Press, Oxford, 1999.

\bibitem{Landau_Binder}
D.~Landau and K.~Binder, {\em A Guide to Monte Carlo Simulations in Statistical
  Physics}.
\newblock Cambridge University Press, New York, NY, USA, 2005.

\bibitem{Ferrenberg_Landau:1991}
A.~M. Ferrenberg and D.~P. Landau, ``Critical behavior of the three-dimensional
  ising model: A high-resolution monte carlo study,'' {\em Physical Review B}
  {\bf 44} (sep, 1991) 5081--5091.

\bibitem{Binder1981}
K.~Binder, ``{Finite size scaling analysis of Ising model block distribution
  functions},'' {\em Z. Phys.} {\bf B43} (1981) 119--140.

\bibitem{Multihisto}
A.~M. Ferrenberg and R.~H. Swendsen, ``Optimized monte carlo data analysis,''
  {\em Phys. Rev. Lett.} {\bf 63} (Sep, 1989) 1195--1198.

\bibitem{Bykov:2011ai}
D.~Bykov, ``{Haldane limits via Lagrangian embeddings},'' {\em Nucl. Phys.}
  {\bf B855} (2012) 100--127, \href{http://xxx.lanl.gov/abs/1104.1419}{{\tt
  1104.1419}}.

\bibitem{Bykov:2012am}
D.~Bykov, ``{The geometry of antiferromagnetic spin chains},'' {\em Commun.
  Math. Phys.} {\bf 322} (2013) 807--834,
  \href{http://xxx.lanl.gov/abs/1206.2777}{{\tt 1206.2777}}.

\bibitem{Lajko:2017wif}
M.~Lajkó, K.~Wamer, F.~Mila, and I.~Affleck, ``{Generalization of the Haldane
  conjecture to SU(3) chains},'' {\em Nucl. Phys.} {\bf B924} (2017) 508--577,
  \href{http://xxx.lanl.gov/abs/1706.06598}{{\tt 1706.06598}}.

\end{thebibliography}\endgroup

\end{document}